\documentclass[aps,prd,floatfix,showpacs,superscriptaddress,nofootinbib,notitlepage]{revtex4-1}

\usepackage{dsfont}

\usepackage{amsmath}

\usepackage{amsfonts}

\usepackage{amssymb}

\usepackage{graphicx}%

\usepackage{subfig}

\usepackage{braket}

\usepackage{float}

\usepackage{slashed}

\usepackage{color}

\usepackage{tensor}

\definecolor{darkgreen}{rgb}{0,0.35,0}
\usepackage[colorlinks=true, pdfstartview=FitV, linkcolor=blue, citecolor=red, urlcolor=magenta]{hyperref}

\def\a{\alpha}

\def\d{\delta}

\def\m{\mu}
\def\n{\nu}

\def\e{\xi}

\def\l{\lambda}
\def\w{\omega}

\def\r{\rho}

\def\Y{\Psi}

\def\vf{\varphi}

\newcommand{\be}{\begin{equation}}
\newcommand{\ee}{\end{equation}}
\newcommand{\ha}{\frac{1}{2}}
\newcommand{\pa}{\partial}

\def\mc{\mathcal}
\newcommand{\beq}{\begin{eqnarray}}
\newcommand{\eeq}{\end{eqnarray}}

\newcommand{\kulak}{KU Leuven Campus Kortrijk---Kulak, Department of Physics, Etienne Sabbelaan 53 bus 7657, 8500 Kortrijk, Belgium}
\newcommand{\ughent}{Ghent University, Department of Physics and Astronomy, Krijgslaan 281-S9, 9000 Gent, Belgium}
\newcommand{\uerj}
{Universidade do Estado do Rio de Janeiro,
	Instituto de F\'isica---Departamento de F\'isica Te\'orica---Rua S\~{a}o Francisco Xavier 524,
	20550-013, Maracan\~{a}, Rio de Janeiro, Brasil}

\begin{document}
	
	\title{Some remarks on the spectral functions of the Abelian Higgs Model}
	\author{D.~Dudal }\email{david.dudal@kuleuven.be}\affiliation{\kulak}\affiliation{\ughent}

	\author{D.~M.~van Egmond }\email{duifjemaria@gmail.com}\affiliation{\uerj}
	
	\author{M.~S.~Guimar\~{a}es}\email{msguimaraes@uerj.br}\affiliation{\uerj}
	\author{O.~Holanda }\email{ozorio.neto@uerj.br}\affiliation{\uerj}
	\author{B.~W.~Mintz }\email{bruno.mintz@uerj.br}\affiliation{\uerj}
	
	\author{L.~F.~Palhares}\email{leticia.palhares@uerj.br}\affiliation{\uerj}
	\author{G.~Peruzzo}\email{gperuzzofisica@gmail.com}\affiliation{\uerj}
	\author{S.~P.~Sorella}\email{silvio.sorella@gmail.com}\affiliation{\uerj}
	\begin{abstract}
		We consider the unitary Abelian Higgs model and investigate its spectral functions at one-loop order. This  analysis allows to disentangle what is physical and what is not at the level of the elementary particle propagators, in conjunction with the Nielsen identities. We highlight the role of the tadpole graphs and the gauge choices to get sensible results.
		We also introduce an Abelian Curci-Ferrari action coupled to a scalar field to model a  massive photon which, like the non-Abelian Curci-Ferarri model,  is left invariant by a modified non-nilpotent BRST symmetry.    We clearly illustrate its non-unitary nature directly from the spectral function viewpoint. This provides a functional analogue of the Ojima observation in the canonical formalism: there are ghost states with nonzero norm in the  BRST-invariant states of the  Curci-Ferrari model.
		
	\end{abstract}
	
	\maketitle
	

	\section{Introduction}
	
	In recent years there has been an increasing interest in the properties of the spectral function (K\"allen-Lehmann density) of two-point correlation functions, especially in non-Abelian gauge theories such as Quantum Chromodynamics (QCD). It was found \cite{bowman2007scaling,Cucchieri:2004mf,Strauss:2012dg,Dudal:2013yva,Dudal:2019gvn}, in lattice simulations for the minimal Landau gauge, that the spectral function of the gluon propagator is not non-negative everywhere, which means that there is no physical interpretation for this propagator like there is for the photon propagator in Quantum Electrodynamics (QED). This behaviour of the gluon spectral function is commonly associated with the concept of confinement \cite{cornwall2013positivity,Krein:1990sf,Roberts:1994dr,Lowdon:2017gpp}. The non-positivity of the spectral function then becomes a reflection of the inability of the gluon to exist as a free physical particle, i.e.~as an observable asymptotic state of the $S$-matrix.
	
	Many properties of the  non-perturbative infrared region of QCD are coherently
	described today by lattice simulations \cite{n1994finster}. Analytically, however, despite the progress made in the last decades, the achievement of a satisfactory understanding of the infrared (IR)  region is still a big challenge. The gluon propagator is not gauge invariant, and therefore one needs to fix a gauge using the Faddeev-Popov (FP) procedure, introducing ghost fields, whilst trading the local gauge invariance with
	the  Becchi-Rouet-Stora-Tyutin (BRST)  symmetry. In the perturbative ultraviolet region of  QCD, the FP gauge fixing procedure  works, giving  results in excellent agreement with experiments. However, an extrapolation of the perturbative results to low energies is plagued by infrared divergencies caused by the existence of the well known Landau pole.
	In the same region, it is known that the standard FP procedure does not fix the gauge uniquely: several  field configurations satisfy the same gauge condition, e.g.~the transverse Landau gauge, leading to the so called Gribov copies \cite{gribov1978quantization,Vandersickel:2012tz}. Over the years, various attempts have been made to deal with this problem in the continuum functional approach, see for example \cite{zwanziger1989local,zwanziger1993renormalizability,Dudal:2008sp,Serreau:2012cg,Capri:2015nzw,Zwanziger:2001kw}.
	
	The problem of the Landau pole in asymptotically free theories can be provisionally circumvented, with an adequate renormalization scheme,  by the introduction of an effective  infrared gluon mass \cite{tissier2011infrared}. This is in accordance with the lattice data, which show that  the gluon propagator reaches  a finite positive value in the deep IR for  space-time Euclidean dimensions $d > 2$, see e.g.~\cite{Cucchieri:2007md,Bogolubsky:2009dc,Maas:2008ri,Cucchieri:2009zt,Bornyakov:2009ug,Oliveira:2012eh,Duarte:2016iko,Dudal:2018cli,Boucaud:2018xup}. Of course, analytically one would like to recover  the massless character of the FP gauge fixing theory in the perturbative UV region.  The use of such theory which implements an effective mass only in  the IR region was first proposed in \cite{cornwall1982dynamical} based on the idea of a momentum-dependent or dynamical gluon mass \cite{Parisi:1980jy,Bernard:1981pg}. For this, the Schwinger-Dyson equations are employed in order to get a suitable gap equation  that governs the evolution of the dynamical gluon mass $m(p^2)$, which vanishes for $p^2\to\infty$. This setup preserves both renormalizability and gauge invariance.  Though, the Schwinger-Dyson equations are an infinite set of coupled equations which require a truncation procedure, see for example \cite{papavassiliou2013effective,Aguilar:2014tka, Cyrol:2016tym,Huber:2018ned,Boucaud:2011ug} for a detailed presentation of the subject.
	
	Recently, a more pragmatic approach was taken in \cite{tissier2010infrared,Serreau:2012cg,Gracey:2019xom}, or in other works like \cite{Comitini:2017zfp,Frasca:2015yva,Siringo:2016jrc}. Instead of a model justified from first principles,  the observations obtained from lattice simulations were used as a guiding principle. The massive gluon propagators observed in lattice simulations for Yang-Mills theories led to considering the following action
	\begin{equation}
	S=\int d^d x\frac{1}{4} F_{\m\n}^a F_{\m\n}^a+ \partial_{\mu} \bar{c}^a (D_{\mu}c)^a+ib^a \partial_{\m}A_{\m}^a+\frac{m^2}{2}A_{\mu}^aA_{\mu}^a
	\label{1}
	\end{equation}
	which is  a Landau gauge FP Euclidean Lagrangian for pure gluodynamics, supplemented with a gluon mass term. This term modifies the  theory in the IR but preserves the FP perturbation theory for momenta $p^2 \gg m^2$. The action \eqref{1} is a particular case of the Curci-Ferrari (CF) model \cite{curci1976class}. The mass term breaks the BRST symmetry of the model,  which turns out to be still  invariant under a modified BRST symmetry \cite{Delduc:1989uc}
	\beq
	s_m A_{\m}^a&=&-D_{\m}^{ab}c^b,\\
	s_m c^a&=& \frac{g}{2}f^{abc}c^b c^c,\\
	s_m \bar{c}^a&=&b^a,\\
	s_m b^a&=&im^2 c^a,
	\label{hallo2}
	\eeq
	which is however not nilpotent since $s_m^2\bar{c}\neq 0$.
	
	The legitimacy of the model \eqref{1}  depends, of course,  on how well it accounts for the lattice results. In \cite{tissier2010infrared,tissier2011infrared,Gracey:2019xom}, it has been shown  that, both at one and two-loop order, the model reproduces quite well  the lattice predictions for  the gluon and ghost propagator. For other applications of \eqref{1},  we refer to  \cite{Pelaez:2014mxa,Reinosa:2014zta,Reinosa:2016iml,Pelaez:2013cpa} . However, from the Kugo-Ojima criterion \cite{kugo1979local}, it is known that nilpotency of the BRST symmetry is indispensable to formulate suitable conditions for the construction of the states of the BRST invariant physical (Fock) sub-space, providing unitarity of the  $S$-matrix. Indeed, in \cite{Ojima:1981fs,deBoer:1995dh} the existence of negative norm states in the $s_m$-invariant subspace (``the  would-be physical subspace'') was confirmed.  Though, it is worth to mention here that the goal of \cite{tissier2010infrared} and follow-up works was \emph{not} to introduce a theory for massive gauge bosons, but to discuss a relatively simple and useful effective description of some non-perturbative aspects of QCD.   Unitarity of the gauge bosons sector  is not so much an issue as one expects them to be undetectable  anyhow, due to confinement. Within this perspective the existence of an exact nilpotent BRST symmetry  becomes a quite relevant property  when trying to generalize the action \eqref{1}  to other gauges than Landau gauge, as next to unitary, one should also expect that the correlation functions of gauge-invariant observables are  gauge-parameter independent. We will come back to this question in a separate work.
	
	In this context, it is worthwhile to investigate the spectral properties
	of massive gauge models to try to shed some light on the infrared behavior of their fundamental fields. The direct comparison with a massive model that preserves the original nilpotent BRST symmetry, such as the Higgs-Yang-Mills model, can be particularly enlightening. In any case, the explicit determination of the spectral properties of
	Higgs theories and the study of the role played by gauge symmetry there is an interesting pursue on its own.
	
	For the spontaneously broken symmetries of the Higgs model, one can fix the gauge by means of 't Hooft $R_{\xi}$-gauge. For the formal limit $\xi \rightarrow \infty$ we end up in the unitary gauge, which is considered the physical gauge as it decouples the non-physical particles. However, this gauge is known to be  non-renormalizable \cite{Peskin:1995ev}. Here, we refer the reader to   \cite{irges2017renormalization} for a recent analysis of the unitary gauge. The Landau gauge, on the other hand, corresponds to $\xi \rightarrow 0$. Most articles on massive Yang-Mills models employ the renormalizable Landau gauge, although it was noticed that this gauge might not be the preferred gauge in non-perturbative calculations \cite{Oehme:1979ai}. In fact, as was recently established in \cite{hayashi2018complex}, the use of the Landau gauge in the massive Yang-Mills model \eqref{1} leads to complex pole masses, which will obstruct a calculation of the K\"all\'{e}n-Lehmann spectral function. Indeed, if at some order in perturbation theory (one-loop as in \cite{hayashi2018complex} for example) a pair of Euclidean complex pole masses appear, at higher order these poles will generate branch points in the complex $p^2$-plane at unwanted locations, i.e.~away from the negative real axis, deep into the complex plane, thereby invalidating a K\"all\'{en}-Lehmann spectral representation. This can be appreciated  by rewriting  the Feynman integrals in terms of Schwinger or Feynman parameters, whose analytic properties can be studied through the Landau equations \cite{Eden:1966dnq}. Let us also refer to ~\cite{Baulieu:2009ha,Windisch:2012sz} for concrete examples.
	
	Understanding the different gauges and their influence on the spectral properties is a delicate subject. This gave us further reason to undertake a systematic study of  the spectral properties of Higgs models. In this paper, we present the results for the simplest case: that of the $U(1)$ Abelian Higgs model. In fact, it turned out that this model is already very illuminating on aspects like positivity of the spectral function, gauge-parameter independence of physical quantities and unitarity. Of course, these properties are not unknown in the Abelian case. This article should therefore not be seen as giving any new information on the physical properties of the Abelian model. Rather, exactly because these properties are so well-known, we are in a better position to understand the problems that we face when calculating  the analytic structure behind some of them within a gauge-fixed setup . This work is therefore a first attempt to understand  analytically  the spectral properties of a Higgs-gauge model  in contrast to those of a non-unitary massive model . As such, it is laying the groundwork  for future work on these properties in the non-Abelian $SU(2)$ Higgs case as well as in the massive model  of \cite{tissier2010infrared}, eq.\eqref{1}, investigating the  origin of the  complex poles structure reported in  \cite{hayashi2018complex,Kondo:2019rpa,Binosi:2019ecz}.
	
	The $U(1)$ Higgs model is known to be unitary \cite{gieres1997symmetries, t1981recent} and renormalizable \cite{Becchi:1974md}. In this work, we consider two propagators: that of the photon, and that of the Higgs scalar field. They are obtained through the calculation of the one-loop corrections to the corresponding $1PI$ two-point functions. After adopting the $R_{\xi}$-gauge, we are left with an exact BRST nilpotent  symmetry.   Of course, the correlation function of BRST invariant quantities should be independent of the gauge parameter. Since the transverse component of the photon propagator is gauge invariant, we should find that the one-loop corrected transverse propagator does not depend on the gauge parameter. As a consequence, the photon pole mass will neither. This property has been proven before by the use of the Nielsen identities, \cite{Haussling:1996rq}, see also \cite{Nielsen:1975fs,Piguet:1984js,gambino2000nielsen}, but never in a direct calculation. The same goes for the Higgs particle propagator: the gauge independence of its pole mass was proven in \cite{Haussling:1996rq}, but never in a direct loop calculation to our knowledge.  We underline here the importance of properly taking into account the tadpole contributions  \cite{Martin:2015lxa,Martin:2015rea}  or, equivalently, the effect on the propagators of quantum corrections of the Higgs vacuum expectation value . Armed with the one-loop results, we are able to calculate the spectral properties of the respective propagators for different values of the gauge parameter. An additional aim of this work is to compare our results with those of a non-unitary massive Abelian model, to clearly pinpoint at the level of spectral functions the differences (and issues) of both unitary and non-unitary massive vector boson models.
	
	This article is organized as follows. In section \ref{s2}, we review the spontaneous symmetry breaking of the $U(1)$ Higgs model and its gauge fixing, as well as the tree-level field propagators and vertices. In Section \ref{jaaa}, we calculate the one-loop propagator of both the photon field and the Higgs field, showing the gauge-parameter independence of the transverse photon propagator and of the Higgs pole mass up to one-loop order. In Section \ref{s4}, we calculate the spectral function of both propagators and in \ref{s5} we compare our results with those of a non-unitary massive Abelian model. We also  address  the residue computation. Section \ref{s6} collects our conclusions and outlook.

	\section{Abelian Higgs model: some essentials \label{s2}}
	We start from the Abelian Higgs classical action with a manifest global $U(1)$ symmetry
	\beq
	S = \int d^4x \left\{\frac{1}{4} F_{\m\n}F_{\m\n} + (D_{\m}\vf)^{\dagger} D_{\m}\vf +\frac{\l}{2}\left(\vf^{\dagger}\vf-\frac{v^2}{2}\right)^2\right\}\label{higgsqed},
	\eeq
	where
	\beq
	F_{\m\n}=\pa_{\m}A_{\n}-\pa_{\n}A_{\m},\nonumber\\
	D_{\m}\vf = \pa_{\m}\vf+ieA_{\m}\vf
	\eeq
	and the parameter $v$ gives the vacuum expectation value (vev) of the scalar field  to first order in $\hbar$ , $\langle \vf \rangle_0 =v$.
	The spontaneous symmetry breaking is implemented by expressing  the scalar field as an expansion around its vev,  namely
	\beq
	\vf=\frac{1}{\sqrt{2}}((v+h)+i\r)\label{higgs},
	\eeq
	where the real part $h$ is identified as  the Higgs field and $\rho$ is the (unphysical) Goldstone boson, with $\langle \rho \rangle=0$.  Here we choose to expand around the classical value of the vev, so that $\langle h \rangle$ is zero at the classical level, but receives loop corrections\footnote{ There is of course an equivalent procedure of fixing $\langle h \rangle$ to zero at all orders, by expanding $\vf$ around the full vev:
		$\vf=\frac{1}{\sqrt{2}}((\langle \vf \rangle+h)+i\r)$. In the Appendix \ref{v} we explicitly show that---as expected---both procedures give the same final results up to a given order.} . The action \eqref{higgsqed} now becomes
	\beq
	S&=&\int d^4 x \,\left\{\frac{1}{4} F_{\m\n}F_{\m\n}+\ha\pa_{\m}h\pa_{\m}h+\ha\pa_{\m}\r\pa_{\m}\r - e\,\r\,\pa_{\m}h\, A_{\m}+e\,(h+v)A_{\m}\pa_{\m}\r\right. \nonumber\\
	&+&\left.\frac{1}{2} e^2 A_{\m}[(h+v)^2 + \r^2]A_{\m}+\frac{1}{8}\l(h^2+2h v +\r^2)^2\right\}\label{fullaction2}
	\eeq
	and we notice that both the gauge field and the Higgs field have acquired the following masses
	\beq
	m^2 = e^2 v^2,\,\, m_{h}^2 = \l v^2.
	\eeq
	With this parametrization, the Higgs coupling $\lambda$ and  the parameter $v$ can be fixed in terms of $m$, $m_h$ and $e$, whose values  will be suitably chosen later on in the text.
	
	Even in the broken phase, the action \eqref{fullaction2} is left invariant by the following gauge transformations
	\beq
	\d A_{\m}&=&-\pa_{\m}\w,\,\,\d \vf = ie\w\vf,\,\,\d\vf^{\dagger}=-ie\w\vf^{\dagger},\nonumber\\
	\d h &=&-e\w\r,\,\, \d \r =e\w(v+h).
	\eeq
	where $\omega$ is the gauge parameter.
	\subsection{Gauge fixing}
	Quantization of the theory \eqref{fullaction2} requires a proper gauge fixing. We shall employ the gauge fixing term
	\beq
	S_{gf}=\int d^4x \left\{\frac{1}{2\xi}\left(\partial_{\m}A_{\m}+\xi m \rho\right)^2\right\},
	\label{34}
	\eeq
	known as the 't Hooft or $R_{\xi}$-gauge, which has the pleasant property of cancelling the mixed term $\int d^4x (ev\;A_{\m}\pa_{\m}\r)$ in the expression \eqref{fullaction2}. Of course, \eqref{34} breaks the gauge invariance of the action. As is well known, the latter is replaced by the BRST invariance. In fact, introducing the FP ghost fields $\bar{c},c$ as well as the  auxiliary field $b$, for the BRST transformations we have
	\beq
	sA_{\m}&=&-\pa_{\m}c,\nonumber\\
	s c&=& 0,\nonumber\\
	s \vf &=& iec\vf ,\nonumber\\
	s\vf^{\dagger} &=& -ie c \vf ^{\dagger},\nonumber\\
	s h &=& -e c \r,\nonumber\\
	s\r &=& e c(v+ h),\nonumber\\
	s\bar{c}&=&ib,\nonumber\\
	s b&=&0.
	\label{brst}
	\eeq
	Importantly, the operator $s$ is nilpotent, i.e.~$s^2=0$, allowing to work with the so-called BRST  cohomology, a useful concept to prove unitarity and renormalizability of the Abelian Higgs model \cite{Becchi:1974md,Becchi:1974xu,kugo1979local}.
	
	We can now introduce the gauge fixing in a BRST invariant way via
	\beq
	\mathcal{S}_{gf}&=&s\int d^d x \left\{-i\frac{\e}{2}\bar{c}b+\bar{c}(\pa_{\m}A_{\m} +\e m \r)\right\},\\
	&=&\int d^d x \left\{\frac{\xi}{2}b^2+ib\pa_{\mu}A_{\m}+i b\xi m \r+\bar{c}\pa^2c-\xi m^2 \bar{c}c- \xi me\bar{c}hc\right\}.
	\eeq
	Notice that the ghosts $(\bar c,c)$ get a gauge parameter dependent mass, while interacting directly  with the Higgs field.
	
	The total gauge fixed BRST invariant action then becomes
	\beq
	S&=&\int d^4 x \,\Bigg\{\frac{1}{4} F_{\m\n}F_{\m\n} +\ha\pa_{\m}h\pa_{\m}h+\ha\pa_{\m}\r\pa_{\m}\r - e\,\r\,\pa_{\m}h\, A_{\m}+ e\,h A_{\m}\pa_{\m}\r+ \frac{1}{2} m^2 A_{\mu}A_{\mu} \nonumber\\
	&+& \frac{1}{2} e^2 A_{\m}[h^2 +2vh+ \r^2]A_{\m}+\frac{1}{8}\l(h^2 +\r^2)(h^2 +\r^2+4h v)+\ha m_h^2 h^2+m A_{\mu}\pa_{\mu}\rho+\frac{\xi}{2}b^2+ib \pa_{\m}A_{\m} \nonumber\\
	&+&ib \xi m \rho +\bar{c}(\pa^2)c - m^2 \e c\bar{c}-m\e e \bar{c}c h\Bigg\}, \label{fullaction}
	\eeq
	with
	\beq
	s S = 0\;. \label{brstinvact}
	\eeq
	
	In Appendix \ref{FR} we collect the propagators and vertices corresponding to the action \eqref{fullaction} of the Abelian Higgs model in the $R_{\xi}$ gauge.

	\section{Photon and Higgs propagators at one-loop \label{jaaa}}
	In this section we obtain the one-loop corrections to the photon propagator, as well as to the propagator of the Higgs boson.  This requires the calculation\footnote{We have used the techniques of modifying integrals into ``master integrals'' with momentum-independent numerators from \cite{passarino1979one}.}, in section \ref{AA} and \ref{hh}, of the Feynman diagrams as shown in Figures \ref{les} and \ref{more}.
	
	Notice that the last four diagrams in Figures \ref{les} and \ref{more} vanish  for $\langle h \rangle=0$. Since we have chosen to expand the $\vf$ field around its classical vev $v$ (cf. \eqref{higgs}), $\langle h \rangle$ has loop contributions which are nonzero and the resulting tadpole diagrams have to be included in the quantum corrections for the propagators\footnote{The diagrams with tadpole balloons are not part of the standard definition of one-particle irreducible diagrams that contribute to the self-energies. However, since the momentum flowing in the vertical $h$-field (dashed) line is zero, they can be effectively included as a momentum-independent term in the self-energies.}.
	
	Of course the final result for the propagators would be the same had we chosen to expand the $\vf$ field around its full vev and required $\langle h\rangle=0$.
	In fact, including the tadpole diagrams in our formulation has the same effect as shifting the masses of the fields to include the one-loop corrections to the Higgs vev $\langle \vf \rangle$, calculated by imposing $\langle h \rangle=0$ (see Appendix \ref{v} for the technical details). These diagrams can actually be seen as a correction to the tree-level mass term: in the spontaneously broken phase the gauge boson mass is given by $m=e\langle \varphi \rangle$, depending thus on $\langle \varphi \rangle$ that receives quantum corrections order by order. Therefore, the full inverse photon propagator can be written as
	\beq
	G_{AA}^{-1}(p^2)&=& p^2 +e^2v^2 + \left(\textrm{1PI diagrams}\right) + \left(\textrm{diagrams with tadpoles}\right)\nonumber\\
	&=& p^2 + e^2\langle \varphi \rangle^2+ \left(\textrm{1PI diagrams}\right)
	\,,
	\eeq
	where the equalities are to be understood up to a given order in perturbation theory and a similar reasoning can be drawn for the Higgs propagator.
	
	In what follows, we shall proceed with the expansion adopted in \eqref{higgs} and include the tadpole diagrams explicitly in our self-energy results.
	
		The calculations are done for arbitrary dimension $d$. In section \ref{d} we will analyze the results for $d=4-\epsilon$, making use of the techniques of dimensional regularization in  the $\overline{\text{MS}}$ scheme.
	\subsection{Corrections to the photon self-energy\label{AA}}
	\begin{figure}[t]
		\includegraphics[width=\textwidth]{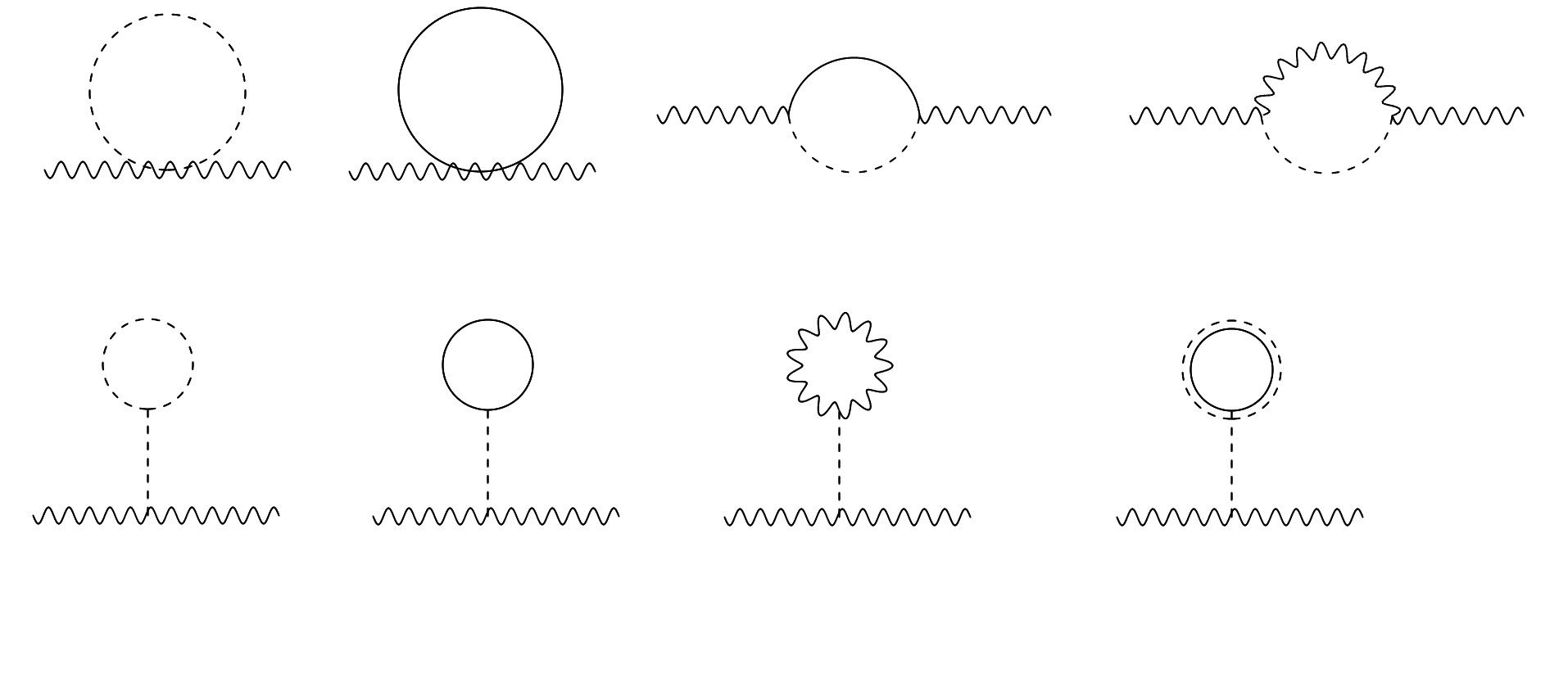}
		\caption{Contributions to one-loop photon self-energy  in the Abelian Higgs Model, including tadpole contributions in the second line. Wavy lines represent the photon field, solid lines the Higgs field, dashed lines the Goldstone boson and double lines the ghost field.}
		\label{les}
	\end{figure}
	The first diagram contributing to the photon self-energy is the Higgs boson snail (first diagram in the first line of Fig. \ref{les}) and gives a contribution
	\beq
	\Gamma_{A_{\m}A_{\n},1}(p^2)&=&\frac{-4e^2}{(4\pi)^{d/2}}\frac{\Gamma(2-d/2)}{2-d}\frac{m_{h}^{d-2}}{2}\delta_{\m\n}.
	\label{besu}
	\eeq
	The second diagram is the Goldstone boson snail (second diagram in the first line of Fig. \ref{les})
	\beq
	\Gamma_{A_{\m}A_{\n},2}(p^2)	&=&\frac{-4e^2}{(4\pi)^{d/2}}\frac{\Gamma(2-d/2)}{2-d}\frac{(\e m^2)^{d/2-1}}{2}\delta_{\m\n}.
	\eeq
	Being momentum-independent, the only effect of these first two diagrams is to renormalize the mass parameters $(m^2_h,m^2)$.
	
	The third term contributing to the photon propagator is the Higgs-Goldstone sunset (third diagram in first line of Fig. \ref{les})
	\beq
	\Gamma_{A_{\m}A_{\n},3}(p^2) &=&\frac{4 e^2}{(4\pi)^{d/2}} \frac{\Gamma(2-d/2)}{2-d}\int_0^1 dx \Bigg[ K_{d/2-1}[m_h^2,\xi m^2]{\mc P}_{\m\n}\nonumber\\
	&+&\left(K_{d/2 -1}[m_h^2,\xi m^2]+\frac{(2-d)}{4}(1-4x(1-x))p^2 K_{d/2 -2}[m_h^2,\xi m^2]\right)\mathcal{L}_{\m\n} \Bigg],
	\eeq

	where we used the definition

	\beq
	K_{\a}(m_1^2,m_2^2)\equiv \Big(p^2x(1-x)+xm_1^2+(1-x)m_2^2\Big)^{\a}.
	\eeq
	The fourth term contributing to the photon propagator is the Higgs-photon sunset (fourth diagram in first line of Fig. \ref{les})
	\beq
		\Gamma_{A_{\m}A_{\n},4}(p^2)
	&=&\frac{4 e^2}{(4\pi)^{d/2}}\frac{\Gamma(2-d/2)}{2-d}\int_{0}^{1} dx \Bigg[\left((2-d)m^2 K_{d/2 -2}(m_h^2,m^2)+K_{d/2-1}(m_h^2,m^2)-K_{d/2 -1}(m_h^2,\xi m^2)\right){\cal P}_{\m\n}\nonumber\\
	&+&\Big((2-d)m^2 K_{d/2 -2}[m_h^2,m^2]+K_{d/2 -1}[m_h^2,m^2]-K_{d/2 -1}[m_h^2,\xi m^2]\nonumber\\
	&&+(2-d)p^2 x^2 (K_{d/2-2}[m_h^2,m^2]-K_{d/2-1}[m_h^2,\xi m^2])\Big)\mathcal{L}_{\m\n}\Bigg].
	\eeq
	
	Finally, we have four tadpole (balloon) diagrams. The Higgs boson balloon (first diagram of the last line in Figure \ref{les})
	\beq
	\Gamma_{A_{\m}A_{\n},5}(p^2)
	&=& \frac{4e^2}{(4\pi)^{d/2}}\frac{\Gamma(2-d/2)}{(2-d)}\frac{3}{2}m_h^{d/2-1}\delta_{\m\n},
	\eeq
	the Goldstone boson balloon (second diagram of the last line in Figure \ref{les})
	\beq
	\Gamma_{A_{\m}A_{\n},6}(p^2)
	&=& \frac{4e^2}{(4\pi)^{d/2}}\frac{\Gamma(2-d/2)}{(2-d)}\frac{1}{2}(\xi m)^{d/2-1}\delta_{\m\n},
	\eeq
	the photon balloon (third diagram of the last line in Figure \ref{les})
	\beq
	\Gamma_{A_{\m}A_{\n},7}(p^2)&=&2e^2 \frac{m^2}{m_h^2}\int \frac{d^dk}{(2\pi)^d}\left( \frac{1}{k^2+m^2}(d-1)+\frac{\xi}{k^2+\xi m^2}\right)\delta_{\m\n},
	\eeq
	and finally, the ghost balloon (fourth diagram of the last line in Figure \ref{les})
	\beq
	\Gamma_{A_{\m}A_{\n},8}(p^2)&=&-2e^2 \frac{m^2}{m_h^2}\int \frac{d^dk}{(2\pi)^d}\frac{\xi}{k^2+\xi m^2}\delta_{\m\n}.
	\label{absu}
	\eeq
	Combining all these contributions \eqref{besu}-\eqref{absu}, we find
	\beq \label{fotonzelf}
	\Gamma_{A_{\m}A{\n}}(p^2)&=&\frac{4 e^2}{(4\pi)^{d/2}}\frac{\Gamma(2-d/2)}{2-d}\int_{0}^{1} dx\left((2-d)m^2 K_{d/2-2}[m^2,m_h^2]+K_{d/2-1}[m^2,m_h^2]+m_{h}^{d-2}+\frac{m^d}{m_h^2}(d-1)\right)P_{\m\n}\nonumber\\
	&+&\frac{4 e^2}{(4\pi)^{d/2}}\frac{\Gamma(2-d/2)}{2-d}\int_{0}^{1} dx\Bigg(\frac{2-d}{4}(1-4x)p^2 K_{d/2 -2}[m_h^2,\xi m^2]+(2-d)(m^2+p^2 x^2)K_{d/2 -2}[m_h^2,m^2]\nonumber\\
	&+&K_{d/2 -1}[m_h^2, m^2]+m_{h}^{d-2}+\frac{m^d}{m_h^2}(d-1)\Bigg)\mathcal{L}_{\m\n}.
	\eeq
	Defining
	\beq
	\Gamma_{A_\m A_\n}=\Pi^{\perp}_{AA}(p^2)\mathcal{P}_{\m\n}+\Pi^{\parallel}_{AA}(p^2) \mathcal{L}_{\m\n},
	\eeq
	it follows that
	\beq
	\partial_{\xi}\Pi^{\perp}_{AA}=0. \label{xiperpind}
	\eeq
	As expected, eq.\eqref{xiperpind}  expresses the gauge parameter independence of the gauge invariant transverse component of the photon propagator \cite{Haussling:1996rq}. Then, the connected transverse form factor,
	\beq
	G_{AA}^\perp(p^2)&=&\frac{1}{p^2+m^2}+\frac{1}{(p^2+m^2)^2}\Pi^{\perp}_{AA}(p^2)+\mathcal{O}(e^4),
	\eeq
	can be rewritten in terms of the resummed form factor as
	\beq
	G_{AA}^\perp(p^2)&=& \frac{1}{p^2+m^2-\Pi^{\perp}_{AA}(p^2)+\mathcal{O}(e^4)}.
	\label{totA}
	\eeq

	\subsection{Corrections to the  Higgs self-energy  \label{hh}}
	
	\begin{figure}[t]
		\includegraphics[width=\textwidth]{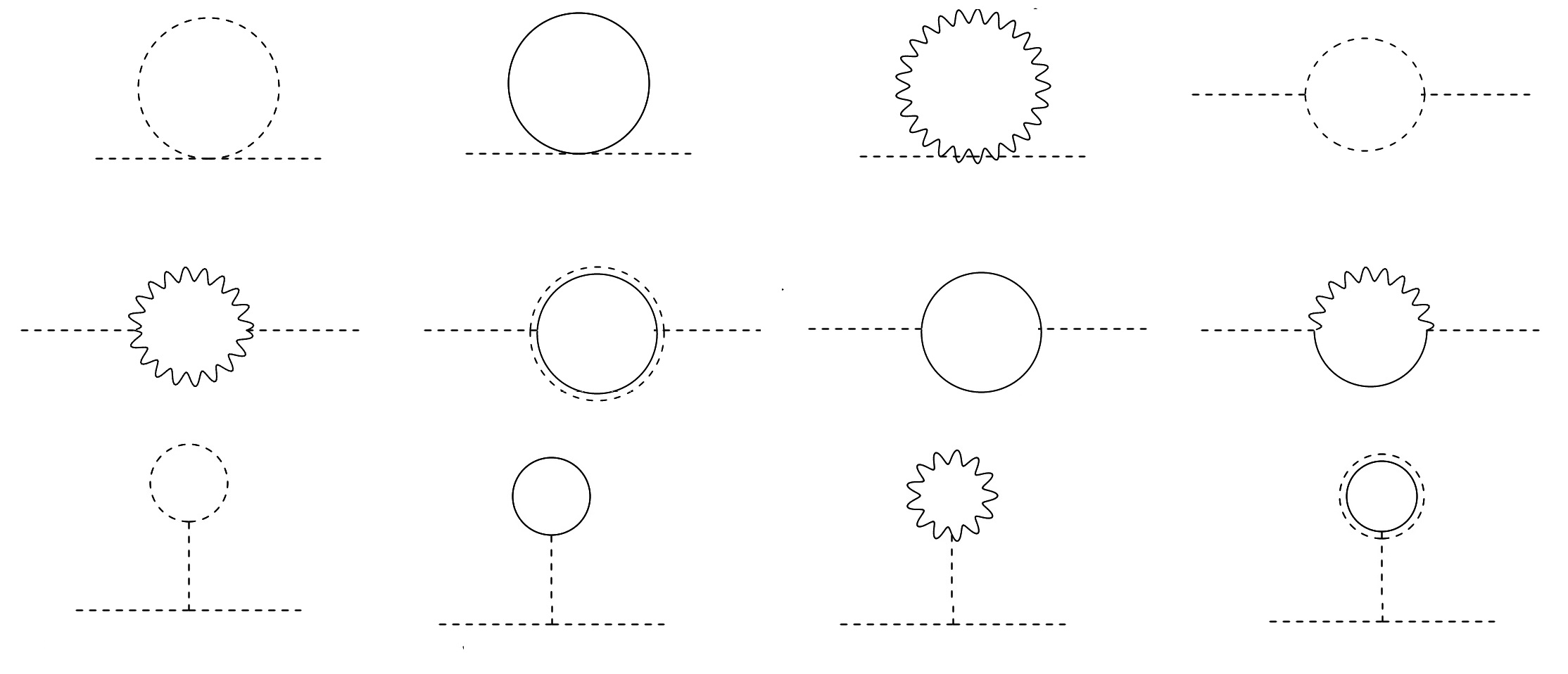}
		\caption{Contributions to the one-loop Higgs self-energy. Line representations are as in Figure 1. }
		\label{more}
	\end{figure}
	The first diagrams contributing to the Higgs self-energy are of the snail type, renormalizing the masses of the internal fields.
	
	The Higgs boson snail (first diagram in the first line of Fig. \ref{more})
	\beq
	\Gamma_{hh,1}(p^2)&=&-3\frac{\lambda}{(4\pi)^{d/2}} \frac{\Gamma(2-d/2)}{(2-d)}m_h^{d-2},
	\label{wat}
	\eeq
	the Goldstone boson snail  (second diagram in the first line of Fig. \ref{more})
	\beq
	\Gamma_{hh,2}(p^2) &=& -\frac{\lambda}{(4\pi)^{d/2}} \frac{\Gamma(2-d/2)}{(2-d)}(\e m^2)^{d/2-1}
	\label{ii}
	\eeq
	and the photon snail  (third diagram in the first line of Fig. \ref{more})
	\beq
	\Gamma_{hh,3}(p^2) &=& -2\frac{e^2}{(4\pi)^{d/2}} \frac{\Gamma(2-d/2)}{(2-d)}\Big( (d-1)m^{d-2}+\e(\e m^2)^{d/2-1}\Big).
	\eeq
	Next, we meet a couple of sunset diagrams. The Higgs boson sunset (fourth diagram in the first line of Fig. \ref{more}):
	\beq
	\Gamma_{hh,4}(p^2)
	&=& \frac{9}{2}\frac{\lambda}{(4\pi)^{d/2}} \frac{\Gamma(2-d/2)}{(2-d)}(2-d)m_h^2 \int_0^1 dx  K_{d/2-2}(m_h^2,m_h^2),
	\eeq
	the photon sunset (first diagram in the second line of Fig. \ref{more}):
	\beq
	\Gamma_{hh,5}(p^2)	&=&  e^2 \frac{\Gamma(2-d/2)}{2-d}\frac{1}{(4\pi)^{d/2}}\int_0^1 dx\Bigg[(2-d)\Big(2m^2(d-1)+2p^2+\frac{p^4}{2m^2}\Big)K_{d/2-2}(m^2,m^2)\nonumber\\
	&-&(2-d)\Big(2p^2+\frac{p^4}{m^2}+\xi^2 m^2+2p^2\xi-2\xi m^2+m^2\Big) K_{d/2-2}(m^2,\xi m^2)\nonumber\\
	&+&(2-d)\Big( 2 \xi p^2+ 2 \xi^2 m^2+ \frac{p^4}{2 m^2}\Big) K_{d/2-2}(\xi m^2, \xi m^2)\nonumber\\
	&+&2(\xi -1) (m^2)^{d/2-1}\nonumber\\
	&+&2(1-\xi)(\xi m^2)^{d/2-1}\Big],
	\eeq
	the ghost sunset (second diagram in the second line of Fig. \ref{more}):
	\beq
	\Gamma_{hh,6}(p^2)
	&=&-\frac{e^2}{(4\pi)^{d/2}} \frac{\Gamma(2-d/2)}{(2-d)} (2-d) m^2 \e^2 \int_0^1 dxK_{d/2-2}(\e m^2, \e m^2),
	\eeq
	the Goldstone boson sunset (third diagram in the second line of Fig. \ref{more}):
	\beq
	\Gamma_{hh,7}(p^2)&=&\frac{1}{2}\frac{\lambda}{(4\pi)^{d/2}} \frac{\Gamma(2-d/2)}{(2-d)}(2-d) m_h^2 \int_0^1 dx K_{d/2-2}(\e m^2, \e m^2)^{d/2-2}
	\eeq
	and a mixed Goldstone-photon sunset (fourth diagram in the second line of Fig. \ref{more}):
	\beq
	\Gamma_{hh,8}(p^2)
	&=& e^2 \frac{\Gamma(2-d/2)}{2-d}\frac{1}{(4\pi)^{d/2}}\int_0^1 dx\Bigg[(2-d)\Big(2p^2+\frac{p^4}{m^2}+\xi^2 m^2+2p^2\xi-2\xi m^2+m^2\Big) K_{d/2-2}(m^2,\xi m^2)\nonumber\\
	&-&(2-d)\Big(\xi^2 m^2+\frac{p^4}{m^2}+2 p^2 \xi \Big)  K_{d/2-2}(\xi m^2,\xi m^2)\nonumber\\
	&+& 2 \Big(1-\xi -\frac{p^2}{m^2}\Big)  (m^2)^{d/2-1}\nonumber\\
	&+& 2\Big( 2\xi - 1+\frac{p^2}{m^2}\Big)(\xi m^2)^{d/2-1}\Bigg].
	\eeq
	Finally, we have the tadpole diagrams. The Higgs balloon (first diagram on the third line of Figure \ref{more}):
	\beq
	\Gamma_{hh,9}(p^2) &=&9 \frac{\lambda}{(4\pi)^{d/2}} \frac{\Gamma(2-d/2)}{(2-d)}m_h^{d-2},
	\eeq
	the photon balloon (second diagram on the third line of Figure \ref{more}):
	\beq
	\Gamma_{hh,10}(p^2) &=& 6\frac{e^2}{(4\pi)^{d/2}} \frac{\Gamma(2-d/2)}{(2-d)}\Big( (d-1)m^{d-2}+\e(\e m^2)^{d/2-1}\Big),
	\eeq
	the Goldstone boson balloon (third diagram on the third line of Figure \ref{more}):
	\beq
	\Gamma_{hh,11}(p^2) &=& 3 \frac{\lambda}{(4\pi)^{d/2}} \frac{\Gamma(2-d/2)}{(2-d)}(\e m^2)^{d/2-1}
	\label{uu}
	\eeq
	the ghost balloon (fourth diagram on the third line of Figure \ref{more}):	
	\beq
	\Gamma_{hh,12}(p^2)&=&-6 \frac{e^2 \e}{(4\pi)^{d/2}} \frac{\Gamma(2-d/2)}{(2-d)} (\e m^2)^{d/2-1}.
	\label{poi}
	\eeq
	Putting together eqs.~\eqref{wat} to \eqref{poi} we find the total one-loop correction to the Higgs boson self-energy,
	\beq
	\Pi_{hh}(p^2)\equiv\Gamma_{hh}(p^2)&=&\frac{\Gamma(2-d/2)}{2-d}\frac{1}{(4\pi)^{d/2}}\int_0^1 dx\Bigg[
	(2-d)e^2\Big(2m^2(d-1)+2p^2+\frac{p^4}{2m^2}\Big)K_{d/2-2}(m^2,m^2)\nonumber\\
	&+& \frac{9}{2}\lambda(2-d)m_h^2 K_{d/2-2}(m_h^2,m_h^2)\nonumber\\
	&+&e^2\left(-2 \frac{p^2}{m^2}+4(d-1)\right) (m^2)^{d/2-1}\nonumber\\
	&+& 6 \lambda (m_h^2)^{d/2-1}\nonumber\\
	&+&(2-d)\Big(-\frac{p^4}{2m^2}e^2+\frac{\lambda}{2}m_h^2\Big) K_{d/2-2}(\xi m^2, \xi m^2)\nonumber\\
	&+& 2(\frac{p^2}{m^2}e^2+\lambda) (\xi m^2)^{d/2-1}\Big].
	\label{pop}
	\eeq
	So, for the Higgs boson resummed connected propagator we find
	\beq
	G_{hh}(p^2)=\frac{1}{p^2+m_h^2-\Pi_{hh}(p^2)}.
	\eeq

	\subsection{Results for $d=4-\epsilon$ \label{d}}
	For $d=4$, the 2-point functions  are divergent. We therefore follow the standard procedure of dimensional regularization, as we have no chiral fermions present. Thus, we choose $d=4-\epsilon$ with $\epsilon$ an infinitesimal parameter, and analyze the solution in the limit $\epsilon \rightarrow 0$.
	
	Let us start with the photon 2-point function, given for arbitrary dimension $d$ by \eqref{fotonzelf}. The mass dimension of the coupling constant $e$ is $[e]=2-d/2=\epsilon/2$, and redefining $e\rightarrow e\tilde{\mu}^{\epsilon/2}=e\tilde{\mu}^{2-d/2}$ we put the dimension on $\tilde{\mu}$, while $e$ is dimensionless. Using
	\beq
	\frac{4 e^2}{(4\pi)^{d/2}}\frac{\Gamma(2-d/2)}{2-d}&\overset{d\rightarrow 4-\epsilon}{=}&
	-2\frac{e^2}{(4\pi)^2} \left(\frac{2}{\epsilon}+1+\text{ln}(\mu^2)\right),
	\eeq
	where we defined
	\beq
	\mu^2=\frac{4\pi \tilde{\mu}^2}{e^{\gamma_E}},
	\eeq we find for the divergent part of the transverse photon 2-point function:
	\beq
	\Pi^\perp_{AA,div}(p^2)
	&=&\frac{2}{\epsilon}\frac{e^2}{(4\pi)^2}\left(\frac{p^2}{3}+6( \frac{g^2}{\lambda}-\frac{1}{2})m^2+3m_h^2\right)
	\eeq
	and these infinities are, following the $\overline{\text{MS}}$-scheme, cancelled by the corresponding counterterms.
	
	The renormalized correlation function is then finite in the limit $d \rightarrow 4$ and we find for the inverse propagator
	\beq
	\frac{1}{G_{AA}^\perp(p^2)}&=&p^2+m^2-2\frac{e^2}{(4\pi)^2}\int_{0}^{1} dx \,\,\Bigg\{K(m^2,m_h^2)(1-\ln\frac{K(m^2,m_h^2)}{\m^2})+m_h^2(1-\ln\frac{m_h^2}{\m^2})\nonumber\\
	&+&\frac{m^4}{m_h^2}(1-3 \ln \frac{m^2}{\m^2})+2m^2 \ln \frac{K(m^2,m_h^2)}{\m^2}\Bigg\},
	\label{dk}
	\eeq
	where we set $K(m_1^2,m_2^2)\equiv K_1(m_1^2,m_2^2)$.
	
	In the same way, we find the divergent part of the Higgs boson 2-point function:
	\beq
	\Pi_{hh,div}(p^2)&=& -\frac{1}{2\epsilon}\frac{1}{(4\pi)^2}\Big(e^2(12p^2-4\xi p^2)+\lambda(8m_h^2-4\xi m^2)\Big),
	\eeq
	which is canceled by the  corresponding counterterm.  Therefore,  the inverse Higgs boson propagator  reads
	\beq
	\frac{1}{G_{hh}(p^2)}&=&p^2+m_h^2+\frac{1}{(4\pi)^2}\int_{0}^{1} dx
	\Bigg\{
	e^2\Big[p^2(1-\ln\frac{m^2}{\m^2}-2\ln \frac{K(m^2,m^2)}{\m^2})\nonumber\\
	&-&\frac{p^4}{2m^2} \ln \frac{K(m^2,m^)]}{\m^2}-6m^2(1-\ln \frac{m^2}{\m^2}+\ln \frac{K(m^2,m^2)}{\m^2})\Big]\nonumber\\
	&+&\lambda \Big[\frac{1}{2}m_h^2(-6+6\ln \frac{m_h^2}{\m^2}-9\ln \frac{K(m_h^2,m_h^2)}{\m^2})\Big]\nonumber\\
	&-&\Big[\xi (e^2p^2+\l m^2)(1- \ln \frac{\xi m^2}{\m^2})-(e^ 2\frac{p^4}{2m^2}-\lambda\frac{m_h^2}{2})\ln \frac{K(\xi m^2,\xi m^2)}{\m^2}\Big]
	\Bigg\}.
	\label{olde}
	\eeq
	Notice that the dependence on the Feynman parameter $x$ in the integrals \eqref{dk} and \eqref{olde} is restricted to functions of the type $\int_0^1 dx \ln \frac{K(m_1^2,m_2^2)}{\mu^2}$. These functions have an analytical solution, depicted in Appendix \ref{apfeyn}.
	
	\section{Spectral properties of the propagators \label{s4}}
	In this section we will  investigate  the spectral properties corresponding to the connected propagators of the last section. Strictly speaking, the calculation of the spectral properties should only be done to  first order\footnote{This would correspond to first order in the gauge coupling $e^2$ and in the Higgs coupling $\lambda$ neglecting the implicit coupling dependence in the masses.} in
	$\hbar$,  since the one-loop corrections to the propagators have been evaluated up to this order.  In practice, however, for small values of the coupling   constants  the higher-order contributions become negligible, and one could treat the one-loop solution as the all-order solution without a significant numerical difference. Even so, when looking for analytical rather than numerical results---for example a gauge parameter dependence---we should restrict ourselves to the first-order results. We shall see the crucial difference between both approaches.
	
	To plot the spectral properties of our model we choose some specific values of the parameters $\{m,m_h,\mu,e\}$. We want to restrict ourselves to the case where the Higgs particle is a stable particle, so we need $m_h^2 < 4m^2$. Furthermore, given the Abelian nature of the model, and thus a weak coupling regime in the infrared, we can choose an energy scale $\mu$ that is sufficiently small w.r.t.~the elusive Landau pole (that is exponentially large) and a corresponding small value for the coupling constant $e$. For the rest of this section and the next, we will therefore choose the parameter values $m=2$  \text{GeV}, $m_h=\frac{1}{2}$ GeV, $\mu=10$ GeV, $e=\frac{1}{10}$.  Notice that by choosing $\mu$ and $e$, we are implicitly fixing the Landau pole $\Lambda$, with $\mu\ll \Lambda$, see \cite{irges2017renormalization} for more details.  We have checked that results are as good as independent from the choice of $\mu$ over a very wide range of $\mu$-values.
	
	We start by calculating the pole mass in section \ref{31}. The pole mass is the actual physical mass of a particle that enters the energy-momentum dispersion relation. It is an observable for both the photon and the Higgs boson and should therefore not depend on the gauge parameter $\xi$. We will also discuss the residue to first order and compare these with the output from the Nielsen identities \cite{Haussling:1996rq}. In section \ref{32} we show how to obtain the spectral function to first order from the propagator. In section \ref{ont} we will discuss some more details about the Higgs spectral function.

	\subsection{Pole mass, residue and Nielsen identities \label{31}}
	The pole mass for any massless or massive field excitation is obtained by calculating the pole of    the resummed connected propagator
	\beq
	G(p^2)= \frac{1}{p^2+m^2-\Pi(p^2)},
	\label{245}
	\eeq
	where $\Pi(p^2)$ is the self-energy correction. The pole of the propagator is thus equivalently defined by the equation
	\beq
	p^2+m^2-\Pi(p^2)=0 \, \label{ppp}
	\eeq
	and its solution defines the pole mass $p^2=-m_{pole}^2$. As consistency requires us to work up to a fixed order in perturbation theory, we should solve eq.\eqref{ppp} for the pole mass in an iterative fashion. To first order in  $\hbar$ , we find
	\beq
	m_{pole}^2=m^2-\Pi^{1-loop}(-m^2)+\mathcal{O}(\hbar^2),
	\label{pol}
	\eeq
	where $\Pi^{1-loop}$ is the first order, or one-loop, correction to the propagator.
	
	Next, we also want to compute the residue $Z$, again up to order   $\hbar$. In principle, the residue is given by
	\beq
	Z= \lim_{p^2 \rightarrow -m_{pole}^2} (p^2+m_{pole}^2)G(p^2).
	\eeq
	We write \eqref{245} in a slightly different way
	\beq
	G(p^2)&=& \frac{1}{p^2+m^2-\Pi(p^2)}\nonumber\\
	&=&\frac{1}{p^2+m^2-\Pi^{1-loop}(-m^2)-(\Pi(p^2)-\Pi^{1-loop}(-m^2))}\nonumber\\
	&=&\frac{1}{p^2+m_{pole}^2-\widetilde{\Pi}(p^2)},
	\label{op}
	\eeq
	where we defined $\widetilde{\Pi}(p^2)=\Pi(p^2)-\Pi^{1-loop}(-m^2)$. At one-loop, expanding  $\widetilde{\Pi}(p^2)$ around $p^2=-m^2_{pole}=-m^2+\mathcal{O}(\hbar)$  gives the residue
	\beq
	Z&=&\frac{1}{1-\partial_{p^2} \Pi(p^2)\vert_{p^2=-m^2}}=1+\partial_{p^2} \Pi(p^2)\vert_{p^2=-m^2}+\mathcal{O}(\hbar^2).
	\label{kak}
	\eeq
	In \cite{Haussling:1996rq}, for the Abelian Higgs model, the Nielsen identities were obtained for both the photon and the Higgs boson. It was found that for the photon propagator, the transverse part is explicitly independent of $\xi$ to all orders of perturbation theory, giving the Nielsen identity:
	\beq
	\partial_{\xi}G^{-1}_{AA}(p^2)=0
	\label{ppo}
	\eeq
	and consequently
	\beq
	\partial_{\xi}\partial_{p^2}G^{-1}_{AA}(p^2) \vert_{p^2=-m_{pole}^2}&=&0,\nonumber\\
	\partial_{\xi}G^{-1}_{AA}(-m_{pole}^2)&=&0,
	\eeq
	confirming the gauge independence of the residue and the pole mass. Of course, this is not unexpected since the transverse part of an Abelian gauge field propagator can be written as
	\begin{equation}
	{\cal P}_{\mu\nu} \braket{A_\mu A_\nu}_{conn} \propto \braket{ A_\mu^T A_\mu^T}\,,\qquad A_\mu^T= {\cal P}_{\mu\nu} A_\nu
	\end{equation}
	and the transverse component $A_\mu^T$  is gauge invariant under Abelian gauge transformations.
	
	We can now compare the outcome of the Nielsen identities with  our one-loop calculation \eqref{dk}. Indeed, to the first order,
	eq.\eqref{dk}  is an explicit demonstration of the identity \eqref{ppo}.
	
	For the Higgs boson, the Nielsen identity is a bit more complicated, and is given by
	\beq
	\partial_{\xi}G^{-1}_{hh}(p^2)=-\partial_{\chi}G^{-1}_{Y_1h}(p^2)G_{hh}^{-1}(p^2),
	\label{sk}
	\eeq
	where  $G^{-1}_{Y_1h}(p^2)$ stands for a non-vanishing $1PI$ Green function which can be obtained from the extended  BRST symmetry which also acts on the gauge parameter \cite{Piguet:1984js}. To be more precise, $Y_1$ is a local source coupled to the BRST variation of the Higgs field (see \eqref{brst}), while $\chi$ is coupled to the integrated  composite operator $\int d^4x\left(-\frac{i}{2}\bar c b+mc\rho\right)$. Acting with $\partial_\chi$ inserts the latter  composite operator with zero momentum flow into the $1PI$ Green function $\braket{(sh) h }$, see \cite{Haussling:1996rq} for the explicit expression of $G^{-1}_{Y_1h}(p^2)$ in terms of Feynman diagrams.
	
\begin{figure}[t]
		\includegraphics[width=10cm]{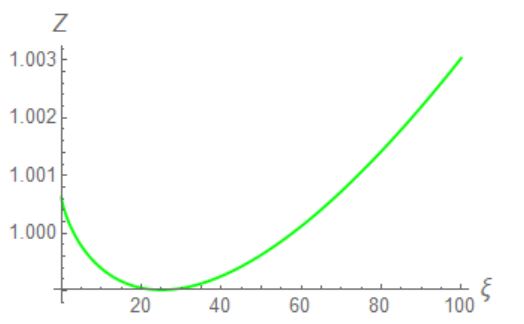}
		\caption{Gauge dependence of the residue of the pole for the Higgs field, for the parameter values $m=2$  \text{GeV}, $m_h=\frac{1}{2}$ GeV, $\mu=10$ GeV, $e=\frac{1}{10}$.}
		\label{Z}
	\end{figure}

	As a consequence, the Higgs propagator $G_{hh}(p^2)$ is not gauge independent, in agreement  with our results \eqref{olde}. From \eqref{sk} we further find
	\beq
	\partial_{\xi}\partial_{p^2}G^{-1}_{hh}(p^2) \vert_{p^2=-m_{pole}^2}&=&-\partial_{\chi}G^{-1}_{Y_1h}(-m_{pole}^2)\partial_{p^2}G_{hh}^{-1}(p^2)\vert_{p^2=-m_{pole}^2},
	\eeq
	which means that the residue is not gauge independent, as $G^{-1}_{Y_1h}(p^2)$ does not necessarily vanish at the pole. We can confirm this for the one-loop calculation, see Figure \ref{Z}.
	Furthermore, we do have
	\beq
	\partial_{\xi}G^{-1}_{hh}(-m_{pole}^2)&=&0,
	\eeq
	so that the Higgs pole mass is indeed gauge independent, the expected result for the physical (observable) Higgs mass. This can be confirmed to one-loop order by using eq.\eqref{pol}, see also Figure \ref{realpole}. Explicitly, in eq.\eqref{olde} for $G^{-1}_{hh}(-m_h^2)$ all the gauge parameter dependence drops out, which means that the Higgs  pole mass is gauge independent to first order in  $\hbar$.

	\subsection{Obtaining the spectral function \label{32}}
\begin{figure}[t]
		\includegraphics[width=10cm]{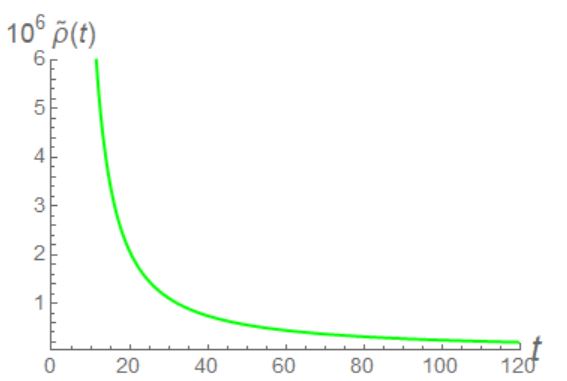}
		\caption{Spectral function of the photon, with $t$ given in $\text{GeV}^2$, for the parameter values $m=2$  \text{GeV}, $m_h=\frac{1}{2}$ GeV, $\mu=10$ GeV, $e=\frac{1}{10}$.}
	\label{photonspectral}
\end{figure}

	We can try to determine the spectral function themselves to first order. To do so, we compare the K\"all\'{e}n-Lehmann spectral representation for the propagator
	\beq
	G(p^2)=\int_0^{\infty} dt \frac{\rho (t)}{t+p^2},
	\label{ltt}
	\eeq
	where $\rho(t)$ is the spectral density function, with the propagator \eqref{op} to first order, written as
	\beq
	G(p^2)&=&\frac{Z}{(p^2+m_{pole}^2-\widetilde{\Pi}(p^2))Z}\nonumber\\
	&=&\frac{Z}{p^2+m_{pole}^2-\widetilde{\Pi}(p^2)+(p^2+m_{pole}^2)\frac{\partial \widetilde{\Pi}(p^2)}{\partial p^2}\vert_{p^2=-m^2}}\nonumber\\
	&=&\frac{Z}{p^2+m_{pole}^2}+Z\left(\frac{\widetilde{\Pi}(p^2)-(p^2+m_{pole}^2)\frac{\partial \widetilde{\Pi}(p^2)}{\partial p^2}\vert_{p^2=-m^2}}{(p^2+m_{pole}^2)^2}\right),
	\label{12}
	\eeq
	where in the last line we used a first-order Taylor expansion so that the propagator has an isolated pole at $p^2=-m_{pole}^2$. In  \eqref{ltt} we can isolate this pole in the same way, by defining the spectral density function as  $\rho(t)=Z \delta(t-m_{pole}^2)+\widetilde{\rho}(t)$, giving
	\beq
	G(p^2)=\frac{Z}{p^2+m_{pole}^2}+ \int_0^{\infty} dt\frac{\widetilde{\rho}(t)}{t+p^2}
	\label{22}
	\eeq
	and we identify the second term in each of the representations \eqref{12} and \eqref{22} as the $\textit{reduced propagator}$
	\beq
	\widetilde{G}(p^2)&\equiv& G(p^2)-\frac{Z}{p^2+m_{pole}^2},
	\eeq
	so that
	\beq
	\widetilde{G}(p^2)= \int_0^{\infty}dt \frac{\widetilde{\rho}(t)}{t+p^2} &=&Z\left(\frac{\widetilde{\Pi}(p^2)-(p^2+m_{pole}^2)\frac{\partial \widetilde{\Pi}(p^2)}{\partial p^2}\vert_{p^2=-m^2}}{(p^2+m_{pole}^2)^2}\right).
	\label{pp}
	\eeq
	Finally, using Cauchy's integral theorem in complex analysis, we can find $\widetilde{\rho}(t)$ as a function of $\widetilde{G}(p^2)$, giving
	\beq
	\widetilde{\rho}(t)=\frac{1}{2\pi i}\lim_{\epsilon\to 0^+}\left(\widetilde{G}(-t-i\epsilon)-\widetilde{G}(-t+i\epsilon)\right).
	\label{key}
	\eeq
	We can now plot the spectral functions for the photon and the Higgs boson.  In Figure \ref{photonspectral} one finds the spectral function for the photon propagator, which is as expected positive definite, next to being $\xi$-independent.  For the record, the threshold (branch point) of the propagator is given by $t^*=2(m_h^2+m^2)$, which can be read off from \eqref{dk}: it corresponds to the smallest value of $-p^2$ where $K(m^2,m_h^2)$ becomes negative.
	
	The Higgs spectral function, on the other hand, is gauge dependent, and therefore it cannot have any direct physical interpretation.
	
\begin{figure}[t]
	\includegraphics[width=10cm]{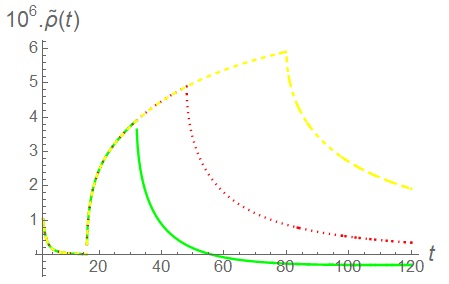}
	\caption{Spectral function of the Higgs boson, with $t$ given in $\text{GeV}^2$, for $\xi=2$ (Green, solid), $\xi=3$ (Red, dotted), $\xi=5$ (Yellow, dashed) and the parameter values $m=2$  \text{GeV}, $m_h=\frac{1}{2}$ GeV, $\mu=10$ GeV, $e=\frac{1}{10}$.}
	\label{higgsspectral}
\end{figure}

	As an illustration, we plot the Higgs spectral function for different values of the gauge parameter in Figure \ref{higgsspectral}. For small values of $t$, the spectral functions for different gauge parameter values are as good as identical, while for larger $t$,  significant differences appear, with the spectral functions for $\xi <3$ even becoming negative . We will relate this to the asymptotic behaviour of the Higgs propagator in the next section, along with some other salient features of the Higgs spectral properties, including its threshold.

\subsection{Some subtleties of the Higgs spectral function \label{ont}}
In this section we  discuss some subtleties that arose during the analysis of the spectral function of the Higgs boson.

\subsubsection{A slightly less correct approximation for the pole mass}	
\begin{figure}[t]
	\includegraphics[width=10cm]{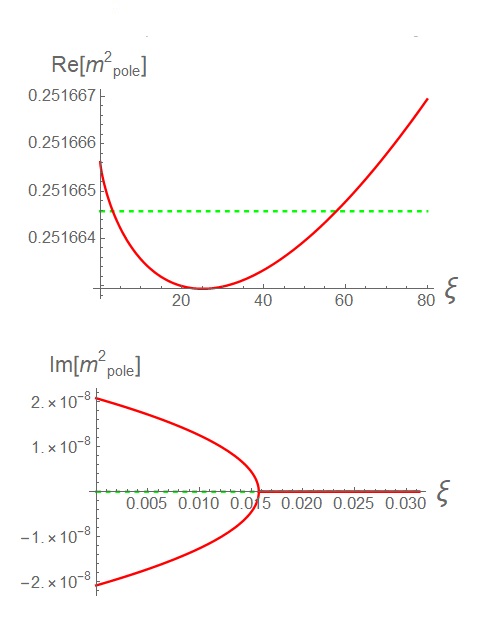}
	\caption{Gauge dependence of the Higgs pole mass obtained iteratively to first order (Green) and the approximated pole mass (Red), for the parameter values $m=2$  \text{GeV}, $m_h=\frac{1}{2}$ GeV, $\mu=10$ GeV, $e=\frac{1}{10}$. Up: real part, down: imaginary part. }
	\label{realpole}
\end{figure}
In the previous two sections we have obtained strictly first-order expressions. In practice, for small values of the coupling parameter $e^2$, we could think about making the approximation
\beq
G(p^2)&=& \frac{1}{p^2+m^2-\Pi(p^2)}\approx  \frac{1}{p^2+m^2-\Pi^{1-loop}(p^2)},
\eeq
in which case one can fix the pole mass by locating the root of
\beq
p^2+m^2-\Pi^{1-loop}(p^2)=0.
\label{ixi}
\eeq
The difference between the pole masses obtained by the iterative method \eqref{pol} and the approximation \eqref{ixi} is very small, of the order $10^{-6}$ for our set of parameters. However, it is rather interesting to notice that the pole mass of the Higgs boson becomes gauge dependent in the approximation \eqref{ixi}. This is no surprise, as the validity of the Nielsen identities is understood either in an exact way, or in a consistent order per order approximation. The previous approximation is neither.

In Figure \ref{realpole} one can see the gauge dependence of the approximated pole mass of the Higgs, in contrast with the first order pole mass. Even worse, for very small values of $\xi$, the approximated pole mass gets complex (conjugate) values. This is due to the fact that the threshold of the branch cut, the branch point, for \eqref{pop} is $\xi$-dependent, as we will see in the next section.

\subsubsection{Something more on the branch points}
\begin{figure}[t]
	\includegraphics[width=10cm]{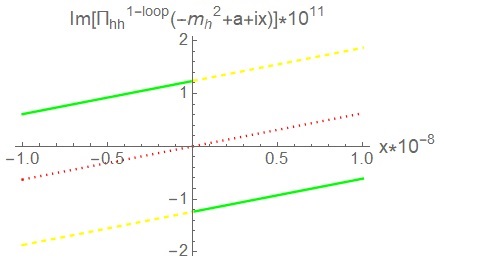}
	\caption{Behaviour of the one-loop correction of the Higgs propagator $\Pi_{hh}(p^2)$ around the pole mass, for the values $a=-10^{-6}$ (Yellow, dashed), $a=0$ (Red, dotted), $a=-10^{-6}$ (Green, solid). The value $x$ is a small imaginary variation of the argument in $\Pi_{hh}(p^2)$.  Only for $a=0$ we find a continuous function at $x=0$, meaning that for any other value, we are on the branch cut.}
	\label{less}
\end{figure}

The existence of a diagram with two internal Goldstone lines (see Figure \ref{more}) leads to a term proportional to $\int_0^1 dx \ln (p^2 x(1-x)+\xi m^2)$ in the Higgs propagator \eqref{pop}. This means that for small values of $\xi$, the threshold for the branch cut of the propagator will be $\xi$-dependent too. Let us look at the Landau gauge $\xi=0$. In this gauge, the above $\ln$-term is proportional to $\ln(p^2)$, due to the now massless Goldstone bosons. This logarithm has a branch point at $p^2=0$, meaning that the pole mass will be lying on the branch cut. Since the first order pole mass is real and gauge independent, this means that
$\Pi_{hh}^{1-loop}(-m_{h}^2)$ is a singular real point on the branch cut. In the slightly less correct approximation of the last section, we will find complex conjugate poles as in Figure \ref{realpole}. This is explained by the fact that for every real value different from $p^2=-m_h^2$, we are on the branch cut, see Figure \ref{less} .  \\

Another consequence of the fact that, for small $\xi$, the pole mass is a real point inside the branch cut is that $\Pi_{hh}(p^2)$ is non-differentiable at $p^2=-m_h^2$ and we cannot extract a residue for this pole. In order to avoid such a problem, we should move away from the Landau gauge and take a larger value for $\xi$, so that the threshold for the branch cut will be smaller than $-m_h^2$. For this we need that $4 \xi m^2 > m_h^2$, which in the case of our parameters set means to require that $\xi > \frac{1}{64}$, in accordance with Figure \ref{realpole}.

\subsubsection{Asymptotics of the spectral function}
Away from the Landau gauge, we see on Figure \ref{higgsspectral} that for e.g.~$\xi=2$ the Higgs spectral function is not non-positive everywhere, while for e.g.~$\xi=5$ it is positive definite, with a turning point at $\xi=3$. How can we explain this difference? The answer can be related to the UV behaviour of the propagator. For $p^2\rightarrow \infty$, the Higgs boson propagator at one-loop behaves as
\beq
G(p^2)=\frac{\mathcal{Z}}{p^2 \ln \frac{p^2}{\m^2}},
\eeq
with $\mathcal{Z}$ depending on the gauge parameter $\xi$. Now, one can show (see Appendix \ref{lot}) that for $\mathcal{Z}>0$, $\rho(t)$ becomes negative for a large value of $t$. For our parameter set, we find that for large momenta
\beq
G^{-1}(p^2)\to (3-\xi)\frac{p^2\ln(p^2)}{1600 \pi^2},\quad\text{for}\quad p^2      \rightarrow \infty,
\eeq
so that for $\xi < 3$, we indeed find $\mathcal{Z}>0$. This indicates that the large momentum behaviour of the propagator makes a difference around $\xi=3$, and determines the positivity of the spectral function, a known fact \cite{Oehme:1979ai,oehme1990superconvergence,Alkofer:2000wg}. This being said, at the same time we cannot trust the propagator values for $p^2 \to\infty$ without taking into account the renormalization group (RG) effects and in particular the running of the coupling, which is problematic for non-asymptotically free gauge theories as the Abelian Higgs model.

\section{A non-unitary $U(1)$ model \label{s5}}
In this section, we will discuss an  Abelian model of the Curci-Ferrari (CF) type  \cite{curci1976slavnov}, in order to compare it  with the Higgs model \eqref{1}. Both models are massive $U(1)$ models with a BRST symmetry. However, while the BRST operator $s$ of the Higgs model is nilpotent, this is not true for the CF-like model. We know the that the Higgs model is unitary but, by the criterion of  \cite{kugo1979local}, the CF model is most probably not.

In section \ref{iiu}  we discuss some essentials for the CF-like model: the action with the modified BRST symmetry, tree-level propagators and vertices. In section \ref{ij} we discuss the one-loop propagators for the photon and scalar field and extract the spectral function. In section \ref{ik} we introduce a local composite field operator that is left invariant by the modified BRST symmetry ot the CF model. The spectral properties of this composite state's propagator will tell us something about the (non-)unitarity of the model, since for unitary models, we expect the propagator of a BRST invariant composite operator to be gauge parameter independent, and the spectral function to be positive definite.

\subsection{CF-like U(1) model: some essentials \label{iiu}}
We start with the action of the CF-like $U(1)$ model
\beq
S_{CF}&=& \int d^d x \left\{\frac{1}{4}F_{\m\n}F_{\m\n}+\frac{m^2}{2}A_{\m}A_{\m}+  (D_{\m}\vf)^{\dagger} D_{\m}\vf +m_{\vf}^2 \vf^{\dagger}\vf+\lambda (\vf \vf^{\dagger})^2\right.\nonumber\\
&-& \left.\a \frac{b^2}{2}+ b\pa_{\mu}A_{\m}+\bar{c} \pa^2 c- \a m^2 \bar{c} c\right\},
\label{loll}
\eeq
where the mass term $\frac{m^2}{2}A_{\m}A_{\m}$ is put in by hand rather than coming from a spontaneous symmetry breaking, and we have fixed the gauge in the linear covariant gauge with gauge parameter $\a$. The mass term breaks the BRST symmetry \eqref{brst} in a soft way. This Abelian CF action is however invariant under the modified BRST symmetry, $s_mS_{CF}=0$, with\footnote{This is the Abelian version of the variation \eqref{hallo2}. For computational purposes, we have also rescaled the $b$-field. Notice that higher order $\alpha$-dependent terms present in the CF model are absent in the Abelian limit \cite{curci1976class,Delduc:1989uc}.}
\beq
s_m A_{\m}&=&-\pa_{\m}c,\nonumber\\
s_m c&=& 0,\nonumber\\
s_m \vf &=& iec\vf ,\nonumber\\
s_m \vf^{\dagger} &=& -ie c \vf ^{\dagger},\nonumber\\
s_m \bar{c}&=&b,\nonumber\\
s_m b&=&-m^2 c.
\label{hallo}
\eeq
As noticed before in our Introduction, this modified BRST symmetry is not nilpotent since $s_m^2 \bar{c} \neq 0$.

From the quadratic part of \eqref{loll} we find the following propagators at tree-level
\beq
\langle A_{\m}(p)A_{\n}(-p)\rangle &=& \frac{1}{p^2+ m^2}{\mc P}_{\m\n}+\frac{\a}{p^2+\a m^2}\mathcal{L}_{\m\n},\nonumber\\
\langle A_{\m}(p)b(-p)\rangle &=& i \frac{p_{\m}}{p^2+\a m^2},\nonumber\\
\langle b(p)b(-p) \rangle &=& -\frac{m^2}{p^2+\a m^2},\nonumber\\
\langle \vf^{\dagger}(p)\vf(-p) \rangle &=& \frac{1}{p^2+m_{\vf}^2},\nonumber\\
\braket{\bar{c}(p)c(-p)}&=&-\frac{1}{p^2+\a m^2},
\eeq
while from the interaction terms we find the vertices
\beq
\Gamma_{A_{\m}\vf^{\dagger}\vf }(-p_1,-p_2,-p_3)&=&e( p_{3,\m}-e p_{2,\m})\delta(p_1+p_2+p_3),\nonumber\\
\Gamma_{A_{\m}A_{\n}\vf^{\dagger}\vf }(-p_1,-p_2,-p_3,-p_4)&=&-2e^2\delta_{\m\n}\delta(p_1+p_2+p_3+p_4),\nonumber\\
\Gamma_{\vf^{\dagger} \vf\vf^{\dagger} \vf}(-p_1,-p_2,-p_3,-p_4)&=& -4\lambda\delta(p_1+p_2+p_3+p_4).
\eeq

\subsection{Propagators and spectral functions \label{ij}}
\begin{figure}[t]
	\includegraphics[width=10cm]{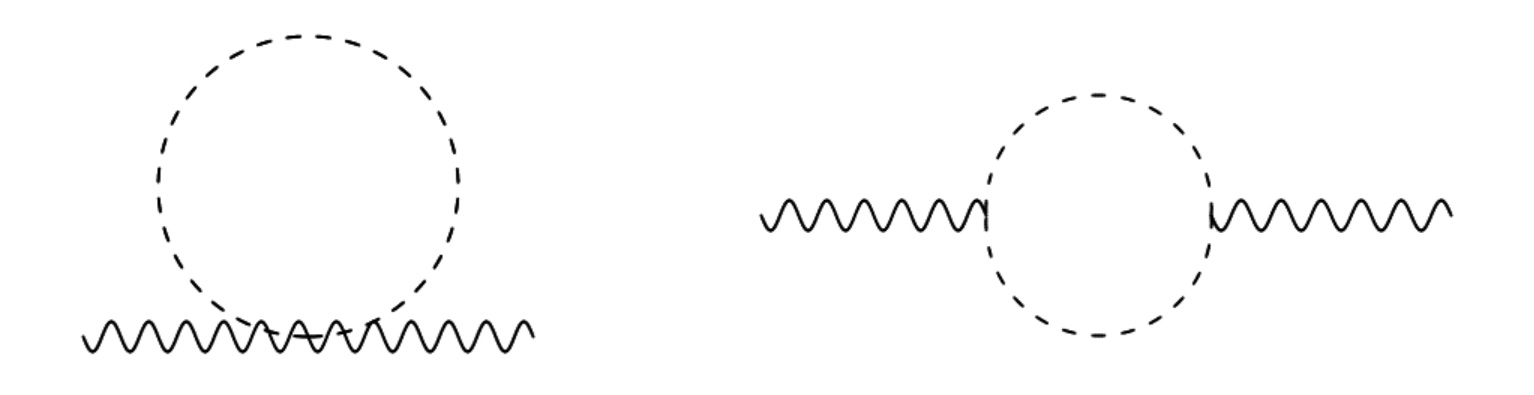}
	\caption{Contributions to one-loop CF photon self-energy.}
	\label{fruit}
\end{figure}
\begin{figure}[t]
	\includegraphics[width=10cm]{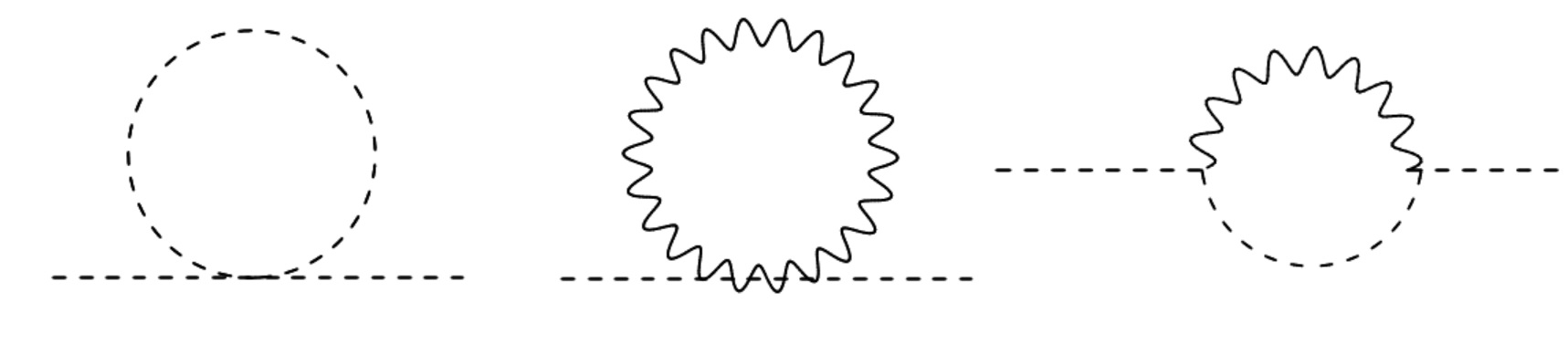}
	\caption{Contributions to the one-loop CF scalar self-energy. Line representations as in Figure 1.}
	\label{7}
\end{figure}
\begin{figure}[t]
\includegraphics[width=10cm]{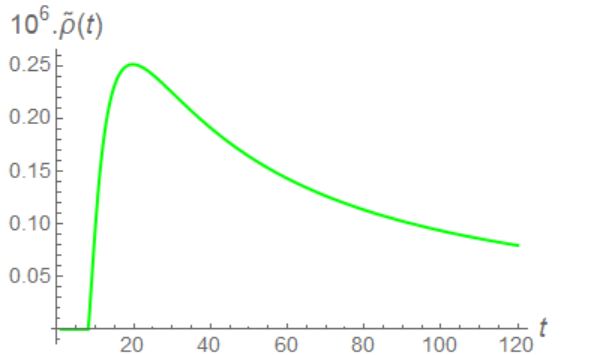}
\caption{Spectral function of the photon field in the Abelian CF model, with $t$ given in $\text{GeV}^2$, for the parameter values $m=2$  \text{GeV}, $m_h=\frac{1}{2}$ GeV, $\mu=10$ GeV, $e=\frac{1}{10}$.}
\label{bbppp}
\end{figure}

The one-loop corrections to the photon and scalar  self-energies  are given in Figure \ref{fruit} and \ref{7}. Without going through the calculational details, we will directly give here the propagators in $d=4$ and discuss some curiosities. The inverse connected photon propagator,
\beq
G^{-1}_{AA}(p^2)&=&p^2+m^2+\frac{e^2}{(4\pi)^2}\int_0^1 dx K(m_{\phi}^2,m_{\phi}^2)(1-\ln \frac{K(m_{\phi}^2,m_{\phi}^2)}{\m^2})-m_{\vf}^2(1-\ln \frac{m_{\vf}^2}{\m^2}),
\label{opop}
\eeq
is independent of the gauge parameter. The threshold of the branch cut is given by $-4 m_{\phi}^2$, and to avoid a pole mass lying on the branch cut, we need to choose here $m^2 < 4m_{\phi}^2$. Choosing as before $m=\frac{1}{2}$  \text{GeV}, $m_{\vf}=2$ GeV, $\mu=10$ GeV, $e=\frac{1}{10}$, we find a positive spectral function, see Figure \ref{bbppp}.

\begin{figure}[t]
\includegraphics[width=10cm]{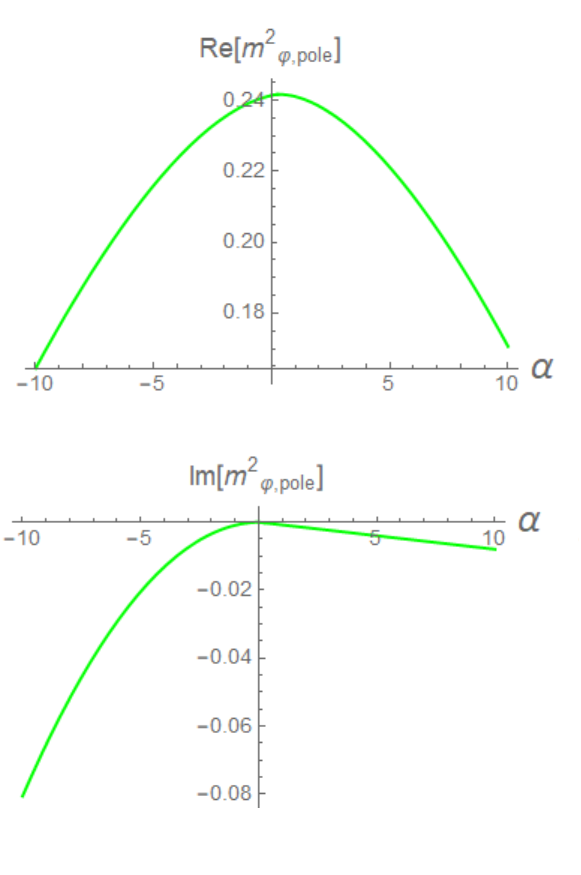}
\caption{Gauge dependence of the first order pole mass for the scalar field. Up: Real part, Down: Imaginary part. The chosen parameter values are $m=\frac{1}{2}$  \text{GeV}, $m_{\vf}=2$ GeV, $\mu=10$ GeV, $e=\frac{1}{10}$.}
\label{23}
\end{figure}

More interestingly, the scalar propagator
\beq
G_{\vf \vf}^{-1}(p)&=&p^2+m_{\vf}^2-\frac{e^2}{(4\pi)^2} \int_0^1 dx \Big(m^2-\a^2 m^2-\a K(m_{\vf}^2,m)(1-2 \ln \frac{K(m_{\vf}^2,\a m^2)}{\m^2})+4K(m_{\vf}^2,0) \frac{p ^2}{m^2}(1-\ln \frac{K(m_{\vf}^2,0)}{\m^2})\nonumber \\
&-& 2K(m_{\vf}^2,m^2) \frac{p^2}{m^2}(1-\ln \frac{K(m_{\vf}^2,m^2)}{\m^2})-2K(m_{\vf}^2,\a m^2) \frac{p^2}{m^2}(1-\ln \frac{K(m_{\vf}^2,\a m^2)}{\m^2})+8 \frac{p^4}{m^2}x^2 \ln \frac{K(m_{\vf}^2,0)}{\m^2}\nonumber \\
&-& 4\frac{p^4}{m^2}x^2 \ln \frac{K(m_{\vf}^2,m^2)}{\m^2}-4 p^2 \ln \frac{K(m_{\vf}^2,m^2)}{\m^2}-(4\a p^2 x - \a p^2 x^2 +4 \frac{p^2}{m^2}x^2)\ln \frac{K(m_{\vf}^2,\a m^2)}{\m^2}-3m^2 \ln \frac{m^2}{\m^2} \nonumber \\
&+& \a m^2 \ln \frac{\a m^2}{\m^2}\Big)+\frac{\l}{(4\pi)^2} m_{\vf}^2(1-\ln \frac{m_{\vf}^2}{\m^2}),
\eeq
is $\a$-dependent, and so is the iterative first-order pole mass $m_{\vf, pole}^2=m_{\vf}^2-\Pi^{1-loop}(-m_{\vf}^2)$. This field can thus not represent a physical particle. For any value other than the Landau gauge $\a=0$ we furthermore get complex poles, see Figure \ref{23}.
From the fact that we find gauge dependent (complex) pole masses for the scalar field, we can already draw the conclusion that the CF model does not describe a physical scalar field. In the next section we will explicitly verify the non-unitary of this model in yet another way.

Essentially, our findings so far mean that in the CF setting, the unphysical gauge parameter $\alpha$ plays here a quite important role,  just like the coupling: different values of the gauge parameter label dfferent theories. This can also be seen from another example: the one-loop vacuum energy of the model will now not only depend on $m$, but also on $\alpha$.

\subsection{Gauge invariant operator \label{ik}}
The Abelian CF model allows us to  construct  a  BRST invariant composite operator $\left(\frac{b^2}{2}+m^2 \bar{c} c \right)$, with
\beq
s_m \left(\frac{b^2}{2}+m^2 \bar{c} c\right)=0.    \label{cop}
\eeq
Although $s_m^2\neq0$  and we can therefore   no longer introduce the BRST cohomology classes, we can still use the fact that $s_m$ is a symmetry generator, thereby defining a would-be physical subspace as the one being annihilated by $s_m$.  A Fock space analogue of this operator was introduced in \cite{Ojima:1981fs},  where  it was established that it has negative norm. As a consequence, it was shown that the physical subspace relating to the symmetry generator $s_m$ was not well-defined, as it contains ghost states. Several more such states were identified later on in \cite{deBoer:1995dh}.

Up to leading order, the connected propagator of the composite operator in eq.\eqref{cop} reads
\beq
G_{\frac{b^2}{2}+m^2 \bar{c} c}(p^2)=\left\langle\left( \frac{b^2}{2}+m^2 \bar{c} c\right) , \left(\frac{b^2}{2}+m^2 \bar{c}c\right) \right\rangle=\frac{1}{4}\langle b^2,b^2\rangle + m^4\langle \bar{c}c, \bar{c}c\rangle,
\label{566}
\eeq

We thus find the propagator \eqref{566} to be
\beq
G_{\frac{b^2}{2}+m^2 \bar{c} c}(p^2)&=&-\frac{3}{4}m^2 \int \frac{d^d k}{(2\pi)^d} \frac{1}{k^2+\a m^2}\frac{1}{(k-p)^2+\a m^2} \nonumber\\
&=&-\frac{3}{4}m^2 \frac{1}{(4\pi)^{d/2}}\Gamma(2-d/2) \int^1_0 dx K_{d/2-2}(\a m^2, \a m^2)
\label{hhu}
\eeq
and this gives for $d=4$, using the $\overline{\text{MS}}$-scheme
\beq
G_{\frac{b^2}{2}+m^2 \bar{c} c}(p^2)&=&\frac{3}{4} \frac{m^2}{(4\pi)^2}\int_0^1 dx \ln \left(\frac{K(\a m^2, \a m^2)}{\m^2}\right).
\eeq
Clearly, the propagator is depending on the gauge parameter $\a$, a not so welcome feature for a presumably physical object.

\begin{figure}[t]
\includegraphics[width=10cm]{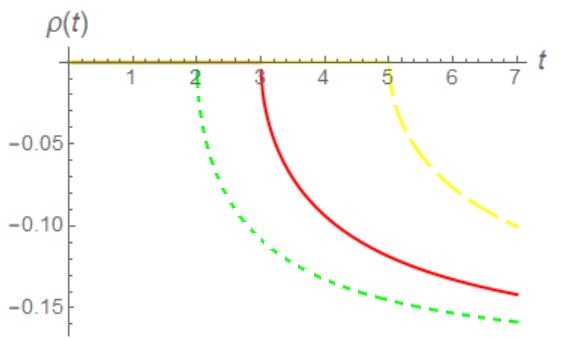}
\caption{Spectral function of the composite operator $\frac{b^2}{2}+m^2 \bar{c} c$, for $\a=2$ (Green, dotted), $\a=3$ (Red, solid), $\a=5$ (Yellow, dashed). The chosen parameter values are $m=\frac{1}{2}$  \text{GeV}, $\mu=10$ GeV.}
\label{bbpp}
\end{figure}

We can also find the spectral function immediately from the propagator by again relying on \eqref{key}. In Figure \ref{bbpp}, one sees that the spectral function is negative for different values of $\a$. Both the $\a$-dependence and the negative-definiteness of the spectral functions demonstrate the non-unitarity of the Abelian CF model. To our knowledge, this is the first time that  ghost-dependent invariant operators  in the physical subspace of a CF model have been constructed from the functional viewpoint\footnote{A similar result can be checked to hold for the original non-Abelian CF model, by adding a few higher order terms to the here introduced Abelian operator. This means that  a non-Abelian version of the operator \eqref{cop}, invariant under the BRST transformation \eqref{hallo2}, can be written down.}, complementing the (asymptotic) Fock space analyses of \cite{Ojima:1981fs,deBoer:1995dh}.

\section{Conclusion and Outlook \label{s6}  }
In the present work we have studied the K\"all\'{e}n-Lehmann spectral  properties  of the $U(1)$ Abelian Higgs model in the $R_{\xi}$ gauge, and that of a $U(1)$ Curci-Ferrari
(CF) like model.

Our main aim was to disentangle in this analytical, gauge-fixed setup what  is  physical  and  what  is  not  at  the  level  of  the elementary particle propagators, in conjunction with the Nielsen identities. Special attention was given to the role played by gauge (in)dependence of different quantities and by the correct implementation of the results up to a given order in perturbation theory.
In particular, calculating the spectral function for the Higgs propagator in the $U(1)$ model, it became apparent that an unphysical occurrence of complex poles, as well as a gauge-dependent pole mass, are caused by the use of the resummed (approximate) propagator as being exact. Indeed, for small coupling constants, the one-loop correction gives a good approximation of the all-order loop correction, and this is a much used method to find numerical results \cite{tissier2010infrared, tissier2011infrared,hayashi2018complex}. However, for analytical purposes, one should stick to the order at which one has calculated the propagator. As we have illustrated, at least in the $U(1)$ Abelian Higgs model case, one will then find a real and gauge independent pole mass for the Higgs boson, in accordance with what the Nielsen identities dictate \cite{Haussling:1996rq}.

Another issue faced here was the fact that the branch point for the Higgs propagator is $\xi$-dependent, being located at $p^2=0$ for the Landau gauge $\xi=0$. For small values of $\xi$, the pole mass has a real value.  However, its value is located   on the branch cut, making it impossible to define a residue at this point, and therefore a spectral function. This means that in order to formulate a spectral function, we should move away from the Landau gauge.
These issues with unphysical (gauge-variant) thresholds are nothing new, see for example \cite{Binosi:2009qm}. They reinforce  in a natural way the need to work with gauge-invariant field operators to correctly describe the observable excitations of a gauge theory.

For the photon, the (transverse) propagator is gauge independent (even BRST invariant), and consequently so are the pole mass, residue and spectral function. For the Higgs boson, the propagator, residue and spectral function are gauge dependent, while the pole mass is gauge independent, in line with the latter being an observable quantity. Notice that the residue of the two-point function does not need to be gauge independent,
since this does not follow from the Nielsen identities, as we discussed in our main text. Rather, the Nielsen identities can be used to show that  the residues of the pole masses in $S$-matrix elements are gauge independent, that is the residues of the singularities in observable scattering amplitudes, see \cite{Grassi:2000dz,Grassi:2001bz}. These residues can evidently be different per scattering process (and per different mass pole). The fact that the Higgs propagator is gauge dependent is not surprising, given that the Higgs field is not invariant under the Abelian gauge transformation.

In future work, it would be interesting to consider, even in perturbation theory, gauge-invariant operators and study their spectral properties using the same techniques of this paper. If the elementary fields are not gauge invariant (like the Higgs field, but also the gluon field in QCD), these
aforementioned gauge-invariant operators will turn out to be composite in nature. Such an approach has recently been  addressed in \cite{Maas:2017xzh,Maas:2017wzi}, based on the seminal observations of Fr\"ohlich-Morchio-Strocchi \cite{Frohlich:1980gj,Frohlich:1981yi}, in which composite operators with the same global quantum numbers (parity, spin, \ldots) as the elementary particles are constructed.\footnote{ For a recent discussion of the renormalization
properties of higher dimensional gauge invariant operators in
Yang-Mills Higgs models see the recent results by \cite{Binosi:2019ecz}.} These composite states will enable us to access directly  the physical spectrum of the theory.   Moreover, we notice that the spectral properties
and the behaviour in the complex momentum  plane of a (gauge-invariant) composite operator will nontrivially depend on the spectral properties of its gauge-variant constituents. This gives another motivation why it is meaningful to study spectral properties of gauge-variant propagators.  Another nice illustrative example of this interplay is the Bethe-Salpeter study of glueballs in pure gauge theories \cite{Sanchis-Alepuz:2015hma}, based on spectral properties of constituent gluons and ghosts \cite{Strauss:2012dg}.
We further
notice that working with gauge-invariant variables will also evade the abovementioned problem with unphysical (gauge-variant) thresholds. Moreover, this methodology could also shed more light on how the confinement-like and Higgs-like phases are analytically connected in the (coupling, Higgs vev)-diagram in the case of a non-Abelian Higgs field in the fundamental representation, thereby making contact with the lattice predictions of Fradkin-Shenker \cite{Fradkin:1978dv,Caudy:2007sf}.

Concluding, in this work several tools have been worked out to determine spectral properties in perturbation theory. We worked up to first order in
$\hbar$
, but everything can be consistently extended to higher orders.  We paid attention how to avoid problems with complex poles and to the pivotal important role of the Nielsen identities, which are intimately related to the exact nilpotent BRST  invariance of the model. These tools will turn out to be quite useful for forthcoming work on the spectral properties of Higgs-Yang-Mills theories. For these theories, the Nielsen identities are well established \cite{gambino2000nielsen}, with supporting lattice data \cite{maas2014two}, providing thus  a solid foundation to compare any results with.

\section*{Acknowledgments}

The authors would like to thank the Brazilian agencies CNPq and FAPERJ for financial support.
This study was financed in part by the Coordena{\c c}{\~a}o
de Aperfei{\c c}oamento de Pessoal de N{\'i}vel Superior---Brasil (CAPES)---Financial Code 001 (M.N.F.). This paper is
also part of the project INCT-FNA Process No.~464898/2014-5.

\appendix
\section{Propagators and vertices of the Abelian Higgs model in the $R_{\xi}$ gauge\label{FR}}

\subsection{Field propagators}
The quadratic part of the action \eqref{fullaction} in the bosonic sector is given by
\beq
S_{bos}^{quad}&=&\ha\int d^4 x \Big\{A_{\m}(-\d_{\m\n}(\pa^2-m^2)+\pa_{\m}\pa_{\n})A_{\n}-\r \pa^2 \r  -h(\pa^2 - m_{h}^2)h +\bar{c}(\pa^2 -m^2 \e )c\nonumber\\
&+&2ib\pa_{\m}A_{\m}+\xi b^2+2im\xi b  \rho +2m A_{\m}\pa_{\m}\rho \Big\}.
\eeq
Putting this in a matrix form yields
\beq
S^{quad}_{bos}=\ha\int d^4 x \,\Y^{T}_{\m} {\mc O}_{\m\n} \Y_{\n},
\eeq
where
\beq
\Y^{T}_{\m}=\left( {\begin{array}{cccc}
A_{\m} &
b&
\r&
h
\end{array} } \right),\,\, \Y_{\n}=\left( {\begin{array}{cccc}
A_{\n}\\
b\\
\r\\
h
\end{array} } \right),
\eeq
and
\beq
{\mc O}=\left( {\begin{array}{cccc}
(-\d_{\m\n}(\pa^2-m^2)+\pa_{\m}\pa_{\n})&-i\pa_{\mu}&m\pa_{\m}&0\\
i \pa_{\n} &\xi &i m \xi &0\\
-m \partial_{\n} &i m \xi&-\pa^2&0\\
0&0&0&-(\pa^2 - m_{h}^2)
\end{array} } \right),
\eeq
the tree-level field propagators can be read off from the inverse of $\mathcal{O}$, leading to the following expressions
\beq
\langle A_{\m}(p)A_{\n}(-p)\rangle &=& \frac{1}{p^2+ m^2}{\mc P}_{\m\n}+\frac{\e}{p^2 + \e m^2}\mathcal{L}_{\m\n},\nonumber\\
\langle \r(p)\r(-p)\rangle &=&\frac{1}{p^2 +\e m^2},\nonumber\\
\langle h(p)h(-p)\rangle &=&\frac{1}{p^2 + m_h^2},\nonumber\\
\langle A_{\m}(p)b(-p)\rangle &=& \frac{p_{\m}}{p^2+\xi m^2},\nonumber\\
\langle b (p) \rho (-k) \rangle &=& \frac{-i m}{p^2+ \xi m^2},
\eeq
where $\mathcal{P}_{\m\n}=\delta_{\m\n}-\frac{p_{\m}p_{\n}}{p^2}$ and $\mathcal{L}_{\m\n}=\frac{p_{\m}p_{\n}}{p^2}$ are the transversal and longitudinal projectors, respectively. The ghost propagator is
\beq
\langle \bar{c}(p)c(-p)\rangle &=&\frac{1}{p^2 +\e m^2}.
\eeq

\subsection{Vertices}
From the action \eqref{fullaction}, we find the following vertices
\beq
\Gamma_{A_{\m}\r h}(-p_1,-p_2,-p_3)&=&ie(p_{\m,3}-p_{\m,2})\delta(p_1+p_2+p_3),\nonumber\\
\Gamma_{A_{\m}A_{\n}h}(-p_1,-p_2,-p_3)&=& -2e^2 v \delta_{\m\n}\delta(p_1+p_2+p_3),\nonumber\\
\Gamma_{A_{\m}A_{\n}hh}(-p_1,-p_2,-p_3,-p_4)&=&-2e^2 \delta_{\m\n}\delta(p_1+p_2+p_3+p_4),\nonumber\\
\Gamma_{A_{\m}A_{\n}\r\r}(-p_1,-p_2,-p_3,-p_4)&=&-2e^2 \delta_{\m\n}\delta(p_1+p_2+p_3+p_4),\nonumber\\
\Gamma_{hhhh}(-p_1,-p_2,-p_3,-p_4)&=&-3\l \, \delta(p_1+p_2+p_3+p_4),\nonumber \\
\Gamma_{hh\r \r}(-p_1,-p_2,-p_3,-p_4)&=&-\l \, \delta(p_1+p_2+p_3+p_4), \nonumber\\
\Gamma_{\r\r\r\r}(-p_1,-p_2,-p_3,-p_4)&=&-3\l \,\delta(p_1+p_2+p_3+p_4), \nonumber\\
\Gamma_{hhh}(-p_1,-p_2,-p_3)&=&-3\l v \,\delta(p_1+p_2+p_3),\nonumber \\
\Gamma_{h\r\r}(-p_1,-p_2,-p_3)&=&-\l v \,\delta(p_1+p_2+p_3), \nonumber\\
\Gamma_{\bar{c}h c}(-p_1,-p_2,-p_3)&=&-m\e e  \,\delta(p_1+p_2+p_3).
\eeq

\section{  Equivalence between including tadpole diagrams in the self-energies and shifting $\langle \vf \rangle$ \label{v}}
There is yet another way to come to \eqref{pop}. For this, we do not need to include the balloon type tadpoles  in the self-energies, but rather fix the expectation value of the Higgs field $\langle h \rangle =0$ by shifting the vacuum expectation value of the Higgs field to its proper one-loop value. The $h$ field one-point function has the following contributions at one-loop order :

\begin{itemize}
\item the gluon contribution
\beq
-\frac{1}{m_h^2}\frac{2 e^2 v}{(4 \pi)^{d/2}}\frac{\Gamma(2-d/2)}{(2-d)}(m^{d-2}(d-1)+\xi (\xi m^2)^{d/2-1}),
\eeq
\item the Goldstone boson one
\beq
-\frac{1}{m_h^2}\lambda v  \frac{1}{(4\pi)^{d/2}} \frac{\Gamma(2-d/2)}{(2-d)}(\e m^2)^{d/2-1},
\eeq
\item the ghost loop
\beq
2 \frac{1}{m_h^2} \frac{ e^2 v \e}{(4\pi)^{d/2}} \frac{\Gamma(2-d/2)}{(2-d)} (\e m^2)^{d/2-1}
\eeq
\item the Higgs boson one
\beq
-3 \frac{1}{m_h^2} \frac{\lambda v}{(4\pi)^{d/2}} \frac{\Gamma(2-d/2)}{(2-d)}m_h^{d-2},
\eeq
\end{itemize}
Together those four contributions yield
\beq
\Gamma_{\braket{h}}&=&\frac{1}{(4\pi)^{d/2}}\frac{\Gamma(2-d/2)}{(2-d)}\frac{1}{m_h^2}(-2e^2v m^{d-2}(d-1)-\lambda v (\xi m^2)^{d/2-1}-3\lambda v m_h^{d-2}),
\label{ll}
\eeq
that becomes, for $d=4-\epsilon$,
\beq
&=& -\frac{1}{2}\frac{1}{m_h^2}\frac{1}{(4\pi)^2}(\frac{2}{\epsilon}+1+\ln (\m^2))(-2e^2 v m^{2-\epsilon}(3-\epsilon)-\lambda v (\xi m^2) ^{1-\epsilon/2}-3\lambda v m_h^{2-\epsilon})\\
&=& -\frac{1}{2}\frac{1}{m_h^2}\frac{1}{(4\pi)^2}(\frac{2}{\epsilon}+1+\ln (\m^2))(-2 e^2 v m^2(1-\frac{\epsilon}{2}\ln m^2)(3-\epsilon)-\lambda v \xi m^2 (1-\frac{\epsilon}{2}\ln m^2)-3 \lambda v m_h^2(1-\frac{\epsilon}{2}\ln m_h^2)).\nonumber
\nonumber\\
\eeq
We can split this in a divergent part
\beq
\Gamma^{div}_{\braket{h}}= \frac{1}{\epsilon}\frac{1}{m_h^2}(6 e^2m^2 v + 3 m_h^2 v \lambda + \xi m^2 v),
\eeq
which we can cancel with the counterterms, and a finite part that reads
\beq
\Gamma_{\braket{h}}^{fin}= \frac{1}{m_h^2}\frac{e^2}{(4 \pi)^2} v \Big( m^2 (1-3 \ln \frac{m^2}{\m^2})\Big) + \frac{1}{m_h^2}\frac{\lambda}{(4 \pi)^2}\frac{v}{2}\Big(3 m_h^2 (1- \ln \frac{h^2}{\m^2})+\xi m^2 (1- \ln \frac{\xi m^2}{\m^2})\Big).
\label{finite<h>}
\eeq

Now, to see how this reflects on the propagator, we can rewrite our scalar field as
\beq
\vf=\frac{1}{\sqrt{2}}((\langle \vf \rangle +h)+i\r),
\eeq
where the vacuum expectation value of the Higgs field has tree-level and one-loop terms:
\beq
\langle \vf \rangle = v + \hbar v_1.
\eeq
Thus the ``classical'' potential part of the action becomes
\beq
\frac{\l}{2}\left(\vf^{\dagger}\vf-\frac{v^2}{2}\right)^2=\frac{\l}{8}\left(\langle \vf \rangle^2-v^2+2h \langle \vf \rangle +h^2+\rho^2\right)^2
\eeq
and expanding this, we find for the shifted tree level Higgs mass
\beq
m_h^2=\frac{1}{2}\lambda (3\langle \vf\rangle^2-v^2)=\lambda v^2+ 3\hbar \l v v_1,
\eeq
while the photon mass is
\beq
m^2= e^2 \langle \varphi \rangle^2= e^2v^2+2\hbar e^2 v v_1.
\eeq

As now per construction $ \langle h \rangle = 0$, we can fix the one-loop correction\footnote{This procedure is also equivalent to computing $\langle \vf \rangle$ via an effective potential minimization up to the same order.} to the Higgs vev by requiring it to absorb the tadpole contributions:
\beq
v_1+ \Gamma^{fin}_{\braket{h}}=0,
\eeq
thus
\beq
v_1= - \frac{1}{m_h^2}\frac{e^2}{(4 \pi)^2} v \Big( m^2 (1-3 \ln \frac{m^2}{\m^2})\Big) - \frac{1}{m_h^2} \frac{\lambda}{(4 \pi)^2}\frac{v}{2}\Big(3 m_h^2 (1- \ln \frac{h^2}{\m^2})+\xi m^2 (1- \ln \frac{\xi m^2}{\m^2})\Big)\,.
\eeq

Implementing this in the transverse $\langle AA \rangle $-propagator, one gets
\beq
G^\perp_{AA}(p^2) = \frac{1}{p^2 + e^2 (v^2+2\hbar v_1 v ) - \Pi^\perp_{AA}(p^2)},
\label{r}
\eeq
where in the correction $\Pi_{AA}^\perp$, which is already of $\mathcal{O}(\hbar)$, we only  include the $\mathcal{O}(\hbar^0)$ part of $\langle\vf\rangle$, i.e. $v$.

We can now verify the $\xi$-independence of the transverse propagator $\langle AA \rangle$. The $\xi$-dependent part of $\Pi_{AA}^\perp(p^2)$ is
\beq
\Pi_{AA,\xi}^\perp(p^2)=\frac{-2e^2}{(4\pi)^{d/2}}\frac{\Gamma(2-d/2)}{2-d}(\e m^2)^{d/2-1},
\eeq
while we find the $\xi$-dependent part of $v_1$ to be (using \eqref{ll})
\beq
v_{1\xi}=\frac{1}{(4\pi)^{d/2}}\frac{\Gamma(2-d/2)}{(2-d)}\frac{1}{m_h^2}(\lambda v (\xi m^2)^{d/2-1}).
\eeq
In the denominator of \eqref{r} we now easily see that
\beq
2e^2 v_{1\xi} v_0-\Pi_{AA,\xi}^\perp(p^2)= 0,
\eeq
thereby establishing the gauge independence of the transverse photon propagator.

For the Higgs propagator, we similarly find
\beq
G_{hh}(p^2) = \frac{1}{p^2 + \lambda (v^2+3\hbar v_1 v ) - \Pi_{hh}(p^2)}.
\label{rr}
\eeq
Here we observe that the $\xi$-dependent part of $v_1$ has the same effect as the balloon tadpole of the Goldstone boson \eqref{uu}, consequently establishing the gauge parameter independence of the Higgs mass pole.
\section{Feynman integrals \label{apfeyn}}

\beq
 \int_0^1 dx \ln \frac{K[m_1^2,m_2^2]}{\mu^2}&=&\frac{1}{2 p^2}\Bigg\{m_1^2 \ln(\frac{m_2^2}{m_1^2})+m_2^2 \ln(\frac{m_1^2}{m_2^2})+p^2 \ln(\frac{m_1^2 m_2^2}{\mu^4})\nonumber\\
&-&2 \sqrt{-m_1^4+2 m_1^2 m_2^2-2 m_1^2 p^2-m_2^4-2 m_2^2 p^2-p^4}\nonumber\\
&\times&\tan^{-1}\Big[\frac{-m_1^2+m_2^2-p^2}{\sqrt{-m_1^4+2 m_1^2 (m_2^2-p^2)-(m_2^2+p^2)^2}}\Big]\nonumber\\
&+&2 \sqrt{-m_1^4+2 m_1^2 m_2^2-2 m_1^2 p^2-m_2^4-2 m_2^2 p^2-p^4}\nonumber\\
&\times&\tan^{-1}\Big[\frac{-m_1^2+m_2^2+p^2}{\sqrt{-m_1^4+2 m_1^2 (m_2^2-p^2)-(m_2^2+p^2)^2}}\Big]\nonumber\\
&-&4 p^2\Bigg\}
\eeq
\section{Asymptotics of the Higgs propagator \label{lot}}
At one-loop, the Higgs propagator behaves like
\begin{equation}\label{dd1}
G_{hh}(p^2)= \frac{\mathcal{Z}}{p^2 \ln \frac{p^2}{\mu^2}} \qquad\text{for}\qquad p^2\to\infty.
\end{equation}
For $\mathcal{Z}>0$, this can only be compatible with
\begin{equation}\label{dd2}
G_{hh}(p^2)= \int_0^\infty \frac{\rho(t)dt}{t+p^2}
\end{equation}
if the superconvergence relation \cite{oehme1990superconvergence,cornwall2013positivity} $\int dt\rho(t)=0$ holds, which forbids a positive spectral function. Let us support this non-positivity of $\rho(t)$ by using \eqref{dd1} to show that $\rho(t)$ is certainly negative for very large $t$. This argument can also be found in the Appendix of \cite{Dudal:2019gvn}.

Since for a KL representation we have:
\beq
\rho(t)=\frac{1}{2\pi i}\lim_{\epsilon\to 0^+}\left(G(-t-i\epsilon)-G(-t+i\epsilon)\right),
\eeq
we find for $t\to+\infty$ and $\epsilon\to 0^+$,
\begin{eqnarray}
\rho(t) &=& \frac{\mathcal{Z}}{2\pi i}\left[\frac{\left(\ln\frac{-t-i\epsilon}{\mu^2}\right)^{-1}}{-t-i\epsilon}-\frac{\left(\ln\frac{-t+i\epsilon}{\mu^2}\right)^{-1}}{-t+i\epsilon}\right]\nonumber \\
&=&\frac{\mathcal{Z}}{2\pi i t} \left[-\left(\ln\frac{t}{\mu^2}-i\pi\right)^{-1}+\left(\ln\frac{t}{\mu^2}+i\pi\right)^{-1}\right]\nonumber\\
&=&\frac{\mathcal{Z}}{\pi t}\text{Im}\left[\left(\ln\frac{t}{\mu^2}+i\pi\right)^{-1}\right]\nonumber\\
&=&\frac{\mathcal{Z}}{\pi t}\left(\left(\ln\frac{t}{\mu^2}\right)^2+\pi^2\right)^{-1/2}\sin\left(-\arctan\frac{\pi}{\ln\frac{t}{\mu^2}}\right).
\end{eqnarray}
From the latter expression, we can indeed infer that $\rho(t)$ becomes negative for $t$ large. We find
\begin{eqnarray}
\rho(t) &\stackrel{t\to\infty}{=}& -\frac{\mathcal{Z} }{t}\left(\ln\frac{t}{\mu^2}\right)^{-2}<0
\end{eqnarray}
for $\mathcal{Z}>0$, and vice versa for $\mathcal{Z}<0$.

\bibliographystyle{ieeetr}

\bibliography{ref}

\end{document}